\documentclass[lettersize,journal]{IEEEtran}
\usepackage{amsthm}
\usepackage{amsmath,amsfonts}
\usepackage{indentfirst}
\usepackage{algorithmic}
\usepackage{array}
\usepackage[caption=false,font=normalsize,labelfont=sf,textfont=sf]{subfig}
\usepackage{textcomp}
\usepackage{stfloats}
\usepackage{url}
\usepackage{verbatim}
\usepackage{graphicx}
\usepackage{stfloats}
\usepackage{multirow}
\usepackage{multicol}
\usepackage{cite}
\usepackage{xcolor}

\newtheorem{theorem}{Theorem}
\newtheorem{corollary}{Corollary}

\newtheorem{remark}{Remark}

\def\BibTeX{{\rm B\kern-.05em{\sc i\kern-.025em b}\kern-.08em
    T\kern-.1667em\lower.7ex\hbox{E}\kern-.125emX}}

\usepackage{balance}
\begin{document}
\title{Exploring Hannan Limitation for 3D Antenna Array}
\author{Ran Ji,~\IEEEmembership{Graduate Student Member,~IEEE}, Chongwen~Huang,~\IEEEmembership{Member,~IEEE}, Xiaoming Chen,~\IEEEmembership{Senior Member,~IEEE}, Wei E. I. Sha,~\IEEEmembership{Senior Member,~IEEE}, Zhaoyang Zhang,~\IEEEmembership{Senior Member,~IEEE}, \\
Jun Yang, Kun Yang, Chau~Yuen,~\IEEEmembership{Fellow,~IEEE}, and M\'{e}rouane~Debbah,~\IEEEmembership{Fellow,~IEEE}
}

\markboth{}%
{How to Use the IEEEtran \LaTeX \ Templates}

\maketitle

\begin{abstract}
    Hannan Limitation successfully links the directivity characteristics of 2D arrays with the aperture gain limit, providing the radiation efficiency upper limit for large 2D planar antenna arrays. This demonstrates the  inevitable radiation efficiency degradation caused by mutual coupling effects between array elements. However, this limitation is derived based on the assumption of infinitely large 2D arrays, which means that it is not an accurate law for small-size arrays. In this paper, we extend this theory and propose an estimation formula for the radiation efficiency upper limit of finite-sized 2D arrays. Furthermore, we analyze a 3D array structure consisting of two parallel 2D arrays. Specifically, we provide evaluation formulas for the mutual coupling strengths for both infinite and finite size arrays and derive the fundamental efficiency limit of 3D arrays. Moreover, based on the established gain limit of antenna arrays with fixed aperture sizes, we derive the achievable gain limit of finite size 3D arrays. Besides the performance analyses, we also investigate the spatial radiation characteristics of the considered 3D array structure, offering a feasible region for 2D phase settings under a given energy attenuation threshold. Through simulations, we demonstrate the effectiveness of our proposed theories and gain advantages of 3D arrays for better spatial coverage under various scenarios. 
\end{abstract}

\begin{IEEEkeywords}
    3D antenna array, Efficiency limit, Gain Limit, 3D beamforming, Hannan Limitation
\end{IEEEkeywords}

\section{Introduction} 
\par Multiple-input-multiple-output (MIMO) technologies have advanced significantly in 4G and 5G communications, proving to be effective means of enhancing communication rates and reliability, and driving substantial progress in the field. 
However, theoretical analysis in electromagnetic (EM) and antenna theories suggests that the maximum achievable gain of a planar array is only dependent on the array aperture size, expressed as $G = \frac{4\pi}{\lambda^2}A_{p}$, where $A_{p}$ represents the physical area of the aperture and $\lambda$ is the wavelength\cite[Eqn. (12-37)]{Balanis2016}. Although this result is derived for a constant distribution aperture, it is equally applicable to arbitrary 2D arrays. While expanding the aperture size of planar arrays to increase the number of antennas can enhance system performance, practical constraints such as the physical sizes of base stations, wind resistance, and hardware costs impose an upper limit on the size of the antenna array's aperture area. 
Additionally, for large 2D arrays, the directivity during beamforming across different spatial directions follows the variation  $D = D_{\max}\cos\theta$ \cite[Eqn. (6-103)]{Balanis2016}\cite{Elliott1964}, where $\theta$  represents the angle between the beamforming direction and the vector perpendicular to the planar array. The $\cos\theta$ factor accounts for the decrease in directivity due to the decrease in the projected area of the array. 
Therefore, the beamforming gain for edge users is significantly reduced, leading to poorer service qualities. 
Due to the aforementioned problems, it is natural to think if it is possible to further facilitate MIMO communications and augment the proven benefits of multi-antenna technologies.

\subsection{Prior Works}

Several technologies have been developed to improve MIMO communications due to its successful implementations \cite{telatar1999capacity,tse2005fundamentals,PAULRAJ2004}. These include massive MIMO communications \cite{Marzetta2010,Larsson2014,BJORNSON20193,Jakob2013}, reconfigurable intelligent surface (RIS) aided MIMO communications \cite{chongwen2019,wankaitang2020,weili2021,chongwen2020}, and the recently emerged Holographic MIMO (HMIMO) communications \cite{Diboya2022,WeiLi2022,jianchengan2023,Ranji2023,weili2023,APizzo2022}, which is developed from the concept of massive MIMO by integrating an immense (potentially infinite) number of antennas into a compact space \cite{APizzo2022}. Specifically, RIS has been demonstrated to have the capability to improve wireless signal coverage \cite{Nemati2020,ganxu2022}.
Moreover, recent advancements in HMIMO communications can partially address the aforementioned issues due to the new physical properties in HMIMO systems such as EM coupling effects and reconfigurable antenna elements \cite{TieruiGong2024}. These new properties enable improved beamforming performance at large angles through superdirectivity \cite{Ranji2024,LHanSuperdirectiveMultiuser}, thus partially resolving the aforementioned challenges.
Review works \cite{CHuang2020,Emil2024,TieruiGong2024} have provided a comprehensive introduction to the emerging HMIMO wireless communications, particularly highlighting the HMIMO near-field communications, theoretical foundations, hardware architectures, and enabling technologies.
However, although these emerging technologies can partially address the current issues, their primary objectives remain to enhance the overall performance of communication systems. 

\par Another significant challenge faced by emerging technologies in MIMO arrays is that their achievable gain remains constrained by the array aperture area \cite{Migliore2019,yuanshuai2022,WanZhongzhichao2023,Mikki2023,RuifengLi2023,Franceschetti_2017}. 
P. Hannan \cite{PHannan1964} investigated the limits of radiation efficiency for antenna elements in an infinitely large planar array and introduced the concept of embedded element efficiency. The results indicate that even for ideal radiating units without ohmic losses, mutual coupling between elements is still inevitable, leading to an average radiation efficiency of less than 1. 
When considering reducing the spacing between elements to place more antennas within the same aperture area, the radiation efficiency of each antenna element decreases, resulting in a constant total array gain. 
Subsequent works such as \cite{Kildal2015, SYuan2023} have verified through numerical measurements the good approximation of the embedded element efficiency method for large practical 2D arrays and its utility in design and numerical analysis. Wasylkiwskyj \cite{Wasylkiwskyj1973,Wasylkiwskyj1974,Wasylkiwskyj19732,Wasylkiwskyj1968} extended research on radiation efficiency issues for different antenna arrays, including linear arrays, finite excitation arrays, and grating lobe arrays, among others. Furthermore, besides radiation efficiency, other performance indicators such as the degrees of freedom (DoF) and achievable gain of antenna arrays are also constrained by the aperture area \cite{Franceschetti_2017,Migliore2019}.


Consequently, it is logical to consider whether applying 3D antenna arrays can break the gain limitations of planar arrays and improve beamforming capabilities for edge users, thus enhancing the overall performance and full-space coverage capabilities of the antenna array.
Studies \cite{Nuttall2001,Costa2018,Sudipta2017} have investigated the directivity and effective area of 3D antenna arrays with different geometric structures. Specifically, for 3D arrays of standard geometric shapes, approximate analytical expressions have been proposed, and for more general arrangements of arbitrary 3D array structures, more complex integral expressions have been provided. In terms of system performance, \cite{xiaohangsong2015,shahu2018,Ranji2024,yuanshuai2024} have studied the communication performance of 3D arrays in both LOS and NLOS scenarios. Specifically, \cite{xiaohangsong2015,shahu2018} revealed the marginal effects of far-field LOS channels. Ji \textit{et al.}\cite{Ranji2024} utilized the electromagnetic coupling effects of HMIMO arrays combined with the 3D array structure to achieve multiple superdirective beams in space and evaluated the spectral efficiency in NLOS scenarios.
Yuan \textit{et al.} \cite{yuanshuai2024} evaluated the DoF and capacity performances of a fabricated two-layer 3D array in both Rayleigh and 3GPP scenarios.
Furthermore, the use of stacked intelligent hypersurfaces \cite{jianchengan2023,jianchengan2024} for interlayer computation and signal processing tasks also shares some conceptual similarities with 3D antenna arrays. However, to the best of the authors' knowledge, no previous work has derived the performance upper bounds for 3D arrays.

\subsection{Our Contribution}
\par In this paper, we extend the theoretical analysis of Hannan Limitation and explore the performance limits for 3D arrays. Specifically, we first introduce the concept of beam feasible region for 2D uniform arrays and give the efficiency upper bound of finite-sized 2D arrays. Furthermore, we derive the fundamental efﬁciency and gain limit for a two-layer 3D array structure.
Moreover, for 3D antenna arrays, we also investigate their spatial radiation characteristics and present the 3D beam feasible region under a given energy attenuation threshold. The contributions of this paper are listed as follows.
\begin{itemize}
    \item We investigate the efficiency upper bound for finite-dimensional 2D arrays and both the efficiency and gain limit for 3D arrays. Specifically, the extension from infinite to finite-size arrays is achieved by analogizing the relationship between the active reflection coefficients and mutual coupling coefficients of the antenna array to the 2D Discrete Fourier Transform (DFT). Through this process, we successfully express the mutual coupling strength of the antenna array as the sampled values of the active reflection coefficients at a finite number of points.
    Moreover, for 3D arrays, the gain limit is proportional to its average spatial projection area. 
    \item For two-layer 3D arrays, we investigate their spatial radiation characteristics and energy attenuation properties under different phase configurations. We also derive the beam feasible regions of the 3D array under a given energy attenuation coefficient constraint. 
    \item Simulation results are provided to validate our proposed theorems and the spatial radiation characteristic analysis of 3D arrays. Additionally, the results quantify the achievable gain of a typical two-layer 3D array structure compared to a planar 2D array (for example, 37.5\% gain for $2\lambda \times 2\lambda$ aperture size with a 0.75$\lambda$ inter-layer spacing).
\end{itemize}


\par The rest of this paper is organized as follows. In Section II, a brief introduction to the derivation of the 2D array efficiency limit is given. Next, theoretical analyses are carried out in Section III to extend the aforementioned methods to the evaluation of the efficiency for finite-dimensional 2D arrays. In Section IV, efficiency and gain limits for two-layer 3D arrays are provided. After that, the spatial radiation characteristics of two-layer 3D arrays are investigated in Section V, and the corresponding spatial feasible region is also presented. Simulation results are provided in Section VI to validate our theoretical analyses and the performance gain of a practical 3D array. Finally, some remarks and conclusions are given in Section VII.

\section{2D Hannan Limitation}
Hannan \cite{PHannan1964} introduced in 1964 the embedded element efficiency concept that explained the so-called element-gain paradox in planar antenna arrays, i.e., that the array gain is always smaller than the sum of the element gain. 
Subsequent works \cite{Kildal2015, SYuan2023} further demonstrated the effectiveness of this approach by evaluating the directivities and aperture efficiencies of planar arrays composed of open-ended waveguides and dipoles, respectively, across a wide range of element spacings.
Specifically, the theoretical value for the embedded element efficiency theory lies in the fact that for $M \times N$ planar array elements, the array gain can be found as $MN$ times the realized gain of the embedded element when the embedded element efficiency is included in the element realized gain model. Moreover, a similar conclusion in a more general form was summarized in \cite{Hahn2007}.
The most important conclusion of these developments is that for a planar array of identical elements situated in identical array environments (i.e., \textit{idempotent array}), and with the same phase shift in adjacent antenna elements, a necessary amount of mutual coupling must exist among the elements of the array, which can be overall represented by the element efficiency parameter. In the following part of this section, we will provide a brief overview of Hannan's derivation approach and explicitly outline the constraint conditions associated with the aforementioned conclusions. This will serve as a foundation for extending these findings to the finite 2D array and 3D array scenarios in the following sections.

\subsection{Array radiation expression}
\begin{figure}[!ht]
    \vspace{0ex}
    \centering
    \includegraphics[width=3.3in]{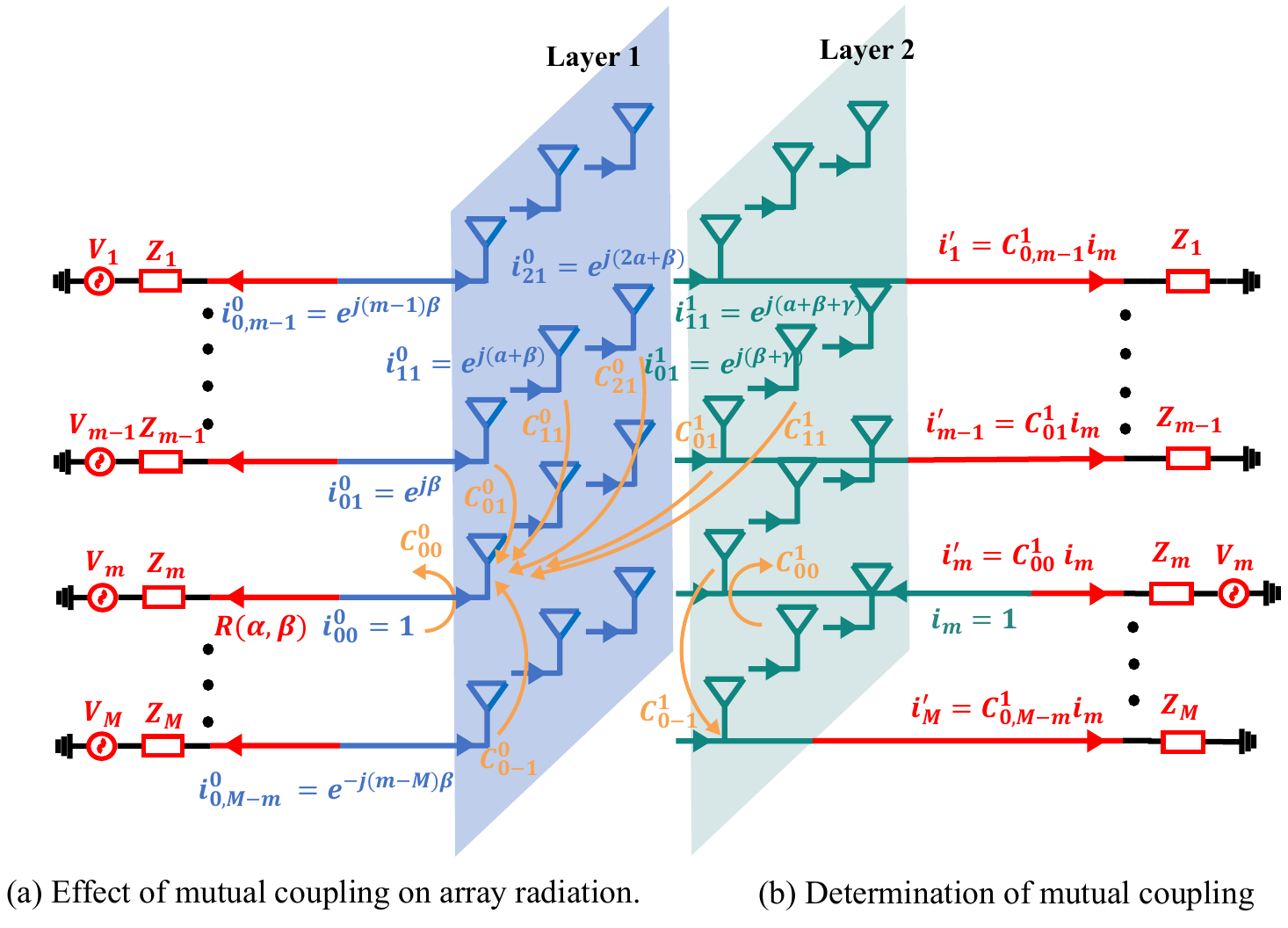}
    \caption{Effect of mutual coupling on array radiation and Determination of mutual coupling.}
    \label{Determination of mutual coupling}
    \vspace{0ex}
\end{figure}

We first consider an infinitely large 2D planar antenna array, as illustrated in Fig. \ref{Determination of mutual coupling}. For 2D cases, only layer 1 exists, and we adopt the concepts of coupling and energy reflection coefficients introduced in \cite{PHannan1964}. The coupling coefficient $C_{pq}$ can be defined through the following process in Fig. \ref{RC2Dinf} (b): by exciting a single element and measuring the reflected energy at all ports, the coupling coefficient is defined as the ratio of the signal reflected at the receiver to the signal available at the generator. This coefficient is generally a complex coupling coefficient, as both the amplitude and the phase are measured. Based on the definition of the coupling coefficient, the energy reflection coefficient can be defined through the symmetrical process shown in Fig. \ref{Determination of mutual coupling} (a): when the entire array is excited with a linear phase difference of $\alpha$ along the $x$-axis and $\beta$ along the $y$-axis, the ratio of the signal reflected at the reference antenna to the signal available at the generator is measured. Finally, the coupling and energy reflection coefficients are related by the following equation due to reciprocity and superposition:
\begin{equation}
    \begin{aligned}
        R(\alpha,\beta) &= \sum_{p=-\infty}^{+\infty}\sum_{q=-\infty}^{+\infty}C_{pq}e^{j\left(p\alpha + q\beta\right)}\\
        &\overset{(a)}{=} \sum_{p=-\infty}^{+\infty}\sum_{q=-\infty}^{+\infty}C_{pq}\cos\left(p\alpha + q\beta\right),
        \label{RC2Dinf}
    \end{aligned}
\end{equation}
which corresponds to the physical interpretation that the reflected signal $R(\alpha,\beta)$ is the sum of the products of the coupling coefficients and the corresponding generator phase factors. Moreover, $C_{pq}$ represents the array coupling coefficients relative to the reference antenna point, where $p$ is the column index and $q$ is the row index, respectively. Equation (a) is due to symmetry $C_{pq}=C_{-p-q}$ in infinitely large arrays.

\par Equation (\ref{RC2Dinf}) is essentially a 2D Fourier transform. Therefore, Parseval's theorem can be used to establish the energy relationship between the reflection coefficients and the coupling coefficients as follows:
\begin{equation}
    \begin{aligned}
        \frac{1}{\pi^2}\int_{\alpha = 0}^{\pi}\int_{\beta = 0}^{\pi}\left|R\left(\alpha,\beta\right)\right|^2 d\alpha d\beta = \sum_{p=-\infty}^{+\infty}\sum_{q=-\infty}^{+\infty}\left|C_{pq}\right|^2,
        \label{P2Dinf}
    \end{aligned}
\end{equation}
which means that \textit{the average power returned to a transmitter in a fully operational array is equal to the net power returned to all transmitters when only one transmitter is excited (the power is averaged over the angle range [0,$\pi$])}.
Moreover, based on the above energy relationship, an expression for the reduction in radiation efficiency due to mutual coupling can be provided:
\begin{equation}
    \begin{aligned}
        \eta &= 1- \sum_{p=-\infty}^{+\infty}\sum_{q=-\infty}^{+\infty}\left|C_{pq}\right|^2 \\
        & = 1- \frac{1}{\pi^2}\int_{\alpha = 0}^{\pi}\int_{\beta = 0}^{\pi}\left|R\left(\alpha,\beta\right)\right|^2 d\alpha d\beta,
    \end{aligned}
\end{equation}

\subsection{2D Feasible Region}
\begin{figure}[!ht]
    \vspace{0ex}
    \centering
    \includegraphics[width=3in]{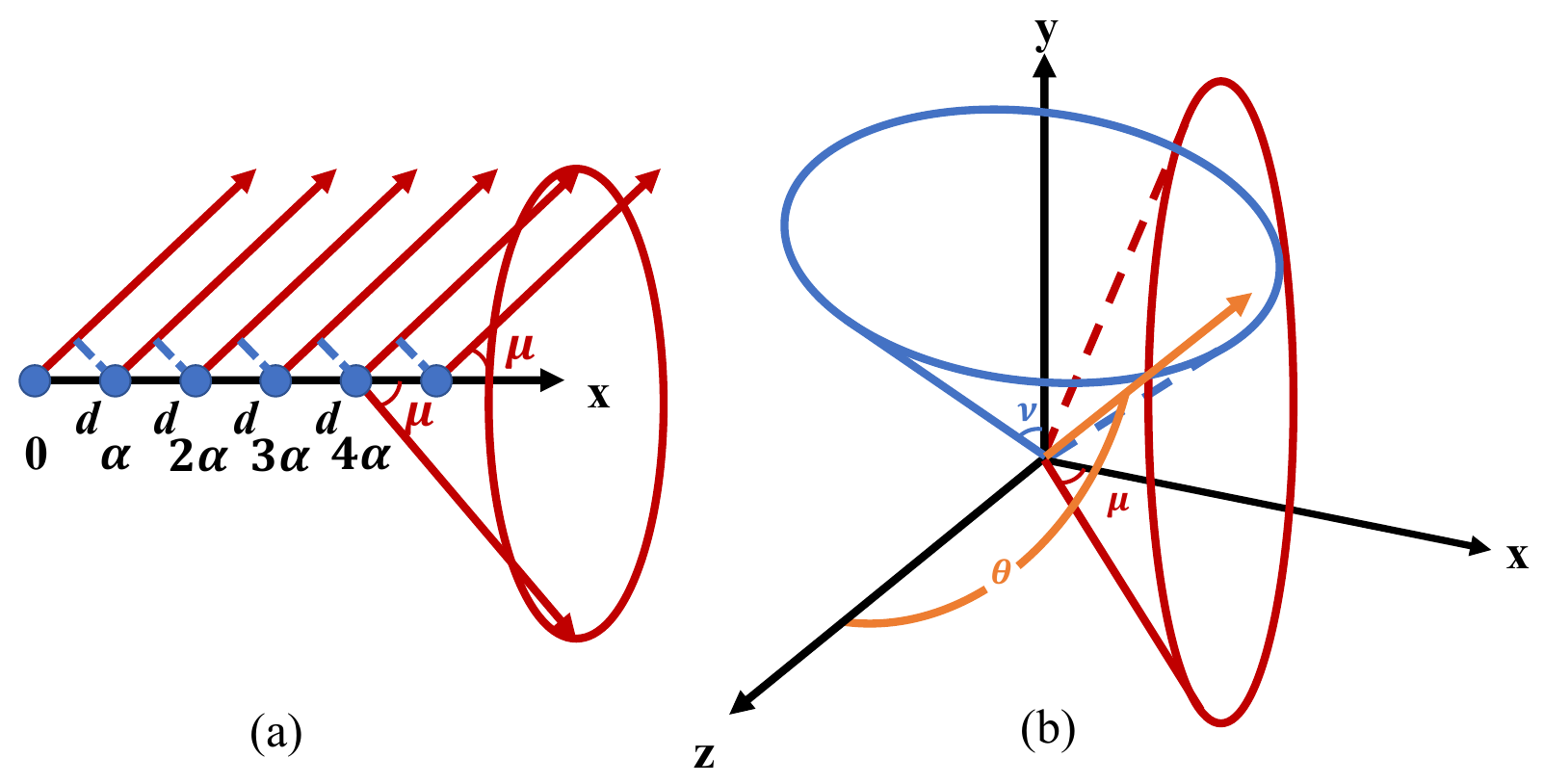}
    \caption{2D array spatial radiation characteristics.}
    \label{2D cone intersection}
    \vspace{-3ex}
\end{figure}

In this subsection, we establish the concept of the beam feasible region for a 2D array and introduce the ideal properties of $R(\alpha,\beta)$ based on this. First, for a linear array along the $x$-axis, the linear phase difference $\alpha$ creates an add-in phase direction in space with an angle $\mu$ to the 
$x$-axis, as shown in Fig. \ref{2D cone intersection} (a). The relationship between them is given by:
\begin{equation}
    \alpha = -2\pi\frac{d_x\cos\mu}{\lambda},
\end{equation}
where $d_x$ is the antenna interspace. Obviously, the direction of phase coherence addition forms a conical shape in space for the linear array. For a 2D array, the phase differences 
$\alpha$ and $\beta$ along the $x$-axis and $y$-axis respectively form two cones with angles $\mu$ and $\nu$ to the $x$-axis and $y$-axis. If these two cones do not intersect, it means the 2D array cannot radiate energy in any direction in space (for an infinitely large array). Conversely, if they do intersect, there exists a spatial radiation direction for the 2D array, as shown in Fig. \ref{2D cone intersection} (b). From the geometric relationship of the figure, it is evident that
$\cos^2\mu+\cos^2\nu = \sin^2\theta$. Therefore, the feasible beam domain of the 2D array can be defined as 
\begin{equation}
    \begin{aligned}
        \cos^2\mu + \cos^2\nu \leq 1 \Rightarrow\left(\frac{\lambda\alpha}{2\pi d_x}\right)^2 + \left(\frac{\lambda\beta}{2\pi d_y}\right)^2 \leq 1,
    \end{aligned}
    \label{2D feasible region}
\end{equation}
Based on the above analysis, a reasonable assumption for the reflection coefficient $R(\alpha,\beta)$ proposed in \cite{PHannan1964} is that $R(\alpha,\beta)=0$ in the feasible region, and $R(\alpha,\beta)=1$ in the non-feasible region where the infinite array cannot effectively radiate. 
The underlying physical reason is as follows: for an infinite 2D array, the spatial radiation beam is infinitely narrow within the feasible region, and its spatial beam shape can be approximately considered equal. 
However, there is no radiation beam in space outside the feasible region 
since coherent phase radiation directions cannot be formed. 
Ultimately, the average radiation efficiency limit of an infinite 2D array is
$1-\frac{1}{\pi^2}\iint_{0}^{\pi}R(\alpha,\beta)d\alpha d\beta = \frac{\pi d_x d_y}{\lambda^2}$ (when there is no grating lobes).
The significance of this radiation efficiency limit expression lies in its connection between the ideal radiation pattern of a 2D planar array (with a 
$\cos\theta$ shape and directivity equals 4) and the maximum achievable gain expression of the antenna array ($\frac{4\pi A_p}{\lambda^2}$). It indicates that the reduction in radiation efficiency due to mutual coupling remains unavoidable.

\section{Radiation Efficiency of 2D Finite Arrays}
\par
\begin{figure}[!ht]
    \vspace{0ex}
    \centering
    \includegraphics[width=2.3in]{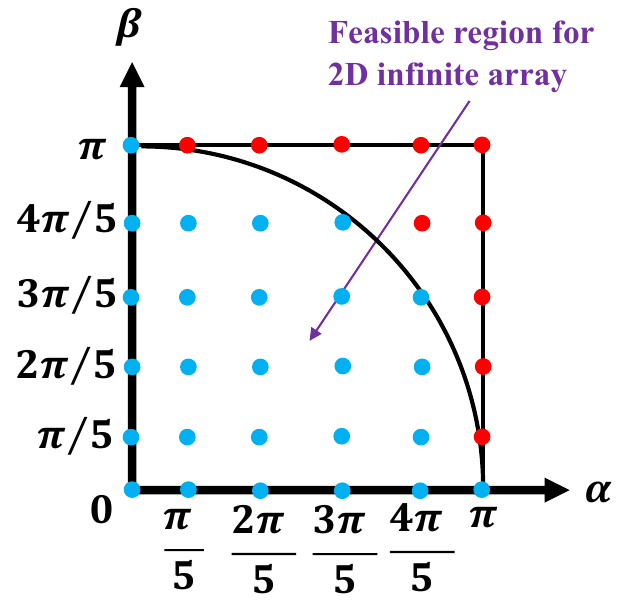}
    \caption{Sampling points for a finite dimension 2D array.}
    \label{2Dsamplingpoints}
    \vspace{0ex}
\end{figure}

 As described in Section II, Hannan established two insightful formulas: the relationship between the coupling and reflection coefficients (\ref{RC2Dinf}), and their corresponding energy relationship (\ref{P2Dinf}),
which are essentially a 2D discrete-time Fourier transform (DTFT) and the corresponding Parseval's theorem. 
In comparison, the standard 2D DTFT transform formulas are shown as follows:
\begin{equation}
    \vspace{-1ex}
    \begin{aligned}
        &X(e^{j\omega_1}, e^{j\omega_2}) = \sum_{m=-\infty}^{\infty} \sum_{n=-\infty}^{\infty} x[m,n] e^{-j(\omega_1 m + \omega_2 n)} \\
        &x[m,n] = \frac{1}{(2\pi)^2} \int_{-\pi}^{\pi} \int_{-\pi}^{\pi} X(e^{j\omega_1}, e^{j\omega_2}) e^{j(\omega_1 m + \omega_2 n)} d\omega_1 d\omega_2,
        \label{2D DTFT}
    \end{aligned}
\end{equation}
By comparing equations (\ref{RC2Dinf}) and (\ref{2D DTFT}), it can be observed that the coupling term  $C_{pq}$  is equivalent to the values of $x[m,n]$ signal. Based on this observation of the above mathematical formulas and the corresponding physical mapping, we can extend the radiation efficiency limits of a 2D infinite planar antenna array to the case of a finite 2D array by analogizing the derivation process from the DTFT to the DFT. This allows us to estimate the radiation efficiency upper limit of a finite-size array. To simplify the analysis, we illustrate this approach using the 1D DTFT and DFT formulas as an example. The 2D DTFT and DFT transformations can be similarly analogized. The formulas for the 1D DTFT and DFT are as follows:
\begin{equation}
    \text{DTFT} \left\{  
        \begin{array}{ll}  
        X(\omega)=\sum_{n=-\infty}^{\infty}x[n]e^{-j\omega n}, &  \\
        \\  
        x[n] = \dfrac{1}{2\pi}\int_{0}^{2\pi}X(\omega)e^{j\omega n}d\omega, &    
        \end{array}
    \right. 
\end{equation}
    
\begin{equation}
\text{DFT}\left\{  
        \begin{array}{ll}  
        X(k)=\sum_{n=0}^{N-1}x[n]e^{-j\frac{2\pi}{N} kn}, &  \\
        \\  
        x[n] = \dfrac{1}{N}\sum_{k=0}^{N-1}X(k)e^{j\frac{2\pi}{N} kn}, &    
        \end{array}
    \right.  
\end{equation}
when the time-domain sampled signal $x[n]$ changes from infinite to finite $N$ points, its frequency-domain characteristics are compressed from being describable at arbitrary points to being described at the sampled points 
$X(\omega=\frac{2\pi k}{N}), k\in\left\{0,1,\hdots,N-1\right\}$. Recall that we consider the time-domain sampled signal $x[n]$ in the equation as equivalent to the coupling terms in the analyzed antenna array scenario, therefore, the finite $N$-point DFT essentially corresponds to the relationship between the coupling and reflection coefficients of a finite-dimensional antenna array. By substituting the time-frequency components in the 2D DFT formula with the 2D antenna array parameters, we can derive the expression for the reflection coefficients of a finite 2D array as follows:
\begin{equation}
    \mathcal{R}_{00}(m,n) = \sum_{p=0}^{M-1}\sum_{q=0}^{N-1}C_{pq}e^{-j\left( \frac{2\pi m}{M}p+\frac{2\pi n}{N}q \right)},
\end{equation}
where $M$ and $N$ represent the number of antenna elements along the $x$-axis and $y$-axis of the 2D array, respectively, and $\mathcal{R}(m,n)$ represents the sampling point of $R(\alpha,\beta)$ at $(\alpha = \frac{2\pi m}{M},\beta = \frac{2\pi n}{N})$. 
The subscript (0,0) indicates that the point at the bottom left of the 2D array in Fig.\ref{Determination of mutual coupling} (a) is selected as the reference point. In this case, the values of the coupling coefficient subscript $(p,q)$ are positive. Clearly, if another point in the array is chosen as the reference point, the range of values for $(p,q)$ will differ. The corresponding new energy relationship is 
\begin{equation}
    \sum_{p=0}^{M-1}\sum_{q=0}^{N-1}\left|C_{pq}\right|^2 = \frac{1}{MN}\sum_{m=0}^{M-1}\sum_{n=0}^{N-1}\left|\mathcal{R}_{00}(m,n)\right|^2,
\end{equation}
However, for finite-dimensional arrays, we essentially focus only on the overall performance of the array. Therefore, by denoting the overall reflection coefficient of the array as $\mathcal{R}(m, n)$, we have the following relationship:
\begin{equation}
    \sum_{p=0}^{M-1}\sum_{q=0}^{N-1}\left|\overline{C}_{pq}\right|^2 = \frac{1}{MN}\sum_{m=0}^{M-1}\sum_{n=0}^{N-1}\left|\mathcal{R}(m,n)\right|^2,
\end{equation}
where $\left|\overline{C}_{pq}\right|$ represent the average coupling strengths of the finite array. 
As an example, we show the sampling points of a $10\times 10$ 2D array in Fig. \ref{2Dsamplingpoints}.

\begin{theorem}
    Element efficiency for a finite 2D array can be formulated as
    \begin{equation}
        \begin{aligned}
            \eta &= 1 - \sum_{p=0}^{M-1}\sum_{q=0}^{N-1}\left|\overline{C}_{pq}\right|^2 \\ 
            & = 1 - \frac{1}{MN}\sum_{m=0}^{M-1}\sum_{n=0}^{N-1}\left|R(\frac{2\pi m}{M},\frac{2\pi n}{N})\right|^2,
        \end{aligned}
        \label{2D finite efficiency equation}
    \end{equation}
    \label{2D finite efficiency}
\end{theorem}
\begin{remark}

    From {\rm \textbf{Theorem \ref{2D finite efficiency}}}, we can transform the problem of estimating the performance limit of a finite 2D antenna array into the problem of estimating the reflected energy at the discrete sampling points, which is further investigated in the following subsections.
\end{remark}

\subsection{Feasible sampling points}

When the discrete sampling points are within the feasible region of the 2D antenna array (as indicated by the blue points in Fig. \ref{2Dsamplingpoints}), the 2D array has a main radiation direction with coherent phase addition in space. Therefore, for a 2D array composed of ideal antenna elements without ohmic losses, we can adopt the assumption consistent with Hannan \cite{PHannan1964}: the reflection coefficient $R$ equals zero inside the feasible region, i.e., $\mathcal{R}(m,n) = 0$.

\subsection{Infeasible sampling points}
\vspace{-0ex}
\par When the discrete sampling points are outside the feasible region (as indicated by the red points in Fig. \ref{2Dsamplingpoints}), the radiation characteristics of the finite 2D array differ from those of an infinite 2D array. 
For a finite array, sampling points outside the feasible region will not result in zero radiated energy in space. Therefore, a correct assessment of the radiation efficiency upper limit for a finite 2D array requires further study of the array reflection coefficients corresponding to sampling points outside the feasible region. 
However, accurately estimating the active reflection coefficient of a finite 2D antenna array under arbitrary excitation conditions is a challenging problem in the fields of electromagnetics and antenna design. In practical scenarios, the commonly used approaches to achieve this are still full-wave electromagnetic simulations or actual measurements, as exemplified by \cite{YangJun2017,Capek2021}. In order to achieve a relatively accurate estimate of the sampling points outside the feasible region without excessively exaggerating the upper limit of the radiation efficiency for finite-size 2D arrays, we adopt the active reflection coefficient model proposed in \cite{Wasylkiwskyj1973}, where only a finite number of antenna elements are excited in an infinite plane. Compared to the finite 2D array under the same settings, this scenario results in more severe ohmic losses and coupling effects for each antenna excitation generator, thereby reducing radiation efficiency. Thus, applying this model represents a more conservative estimate of the upper limit of the radiation efficiency for a finite 2D array.

\par The estimation formula of reflection coefficients $R(\alpha,\beta)$ for $M \times N$ 2D finite excitation elements in \cite{Wasylkiwskyj1973} can be written as
\begin{equation}
    \resizebox{1\hsize}{!}{$
    \left|R^{(MN)}(\alpha,\beta)\right|^2 = \dfrac{1}{(2\pi)^2}\int_{-\pi}^{\pi}\int_{-\pi}^{\pi}\left|R(\varrho,\zeta)\right|^2\left|A^{(MN)}(\varrho,\zeta;\alpha,\beta)\right|^2 d\varrho d\zeta,  
    $}
\end{equation}
where $R(\varrho,\zeta)$ represents the standard reflection coefficient of infinitely large planar arrays (i.e., zero in the feasible region and one in the non-feasible region). Moreover, the excitation function can be defined as 
\begin{equation}
    A^{(MN)}(\varrho,\zeta;\alpha,\beta) = \sum_{m=-\infty}^{+\infty}\sum_{n=-\infty}^{+\infty}a_{m}b_n e^{-j(m\varrho+n\zeta)},
\end{equation}
for a $M\times N$ finite dimension array, we apply 
\begin{equation}
    \begin{aligned}
        a_m &= 
        \begin{cases} 
        \frac{1}{\sqrt{MN}} e^{jm\alpha}, & |m| \leq \frac{M-1}{2} \\
        0, & |m| > \frac{M-1}{2}
        \end{cases} \\
        b_m &= 
        \begin{cases} 
        \frac{1}{\sqrt{MN}} e^{jn\beta}, & |n| \leq \frac{N-1}{2} \\
        0, & |n| > \frac{N-1}{2}
        \end{cases}
    \end{aligned}
\end{equation}
and the excitation function can be written as
\begin{equation}
    \resizebox{0.97\hsize}{!}{$
    \left|A^{(MN)}(\varrho,\zeta;\alpha,\beta)\right| = \frac{\sin\left[(M/2)(\alpha-\varrho)\right]\sin\left[(N/2)(\beta-\zeta)\right]}{\sqrt{MN}\sin\left[(\alpha-\varrho)/2\right]\sin\left[(\beta-\zeta)/2\right]},$}
\end{equation}
where $\left|A^{(MN)}(\varrho,\zeta;\alpha,\beta)\right|$  represents the Fourier transform projection component of the finite excitation sequence $a_m$ and $b_n$ onto the standard infinite uniform excitation $(\varrho,\zeta)$. Essentially, the above formula connects the reflection coefficients of finite and infinite arrays through the projection coefficients. Finally, by substituting the estimated reflection coefficients $R^{(MN)}(\alpha,\beta)$ for the sampling points in the infeasible region in (\ref{2D finite efficiency equation}), we can obtain the efficiency upper bound for finite 2D antenna arrays. In the next section, we extend the analysis results for 2D antenna arrays to 3D cases and provide the efficiency and gain limits for 3D arrays.

\section{Performance Analyses of 3D Arrays}
\subsection{Efficiency Analyses}
\begin{figure}[!ht]
    \vspace{0ex}
    \centering
    \includegraphics[width=2.4in]{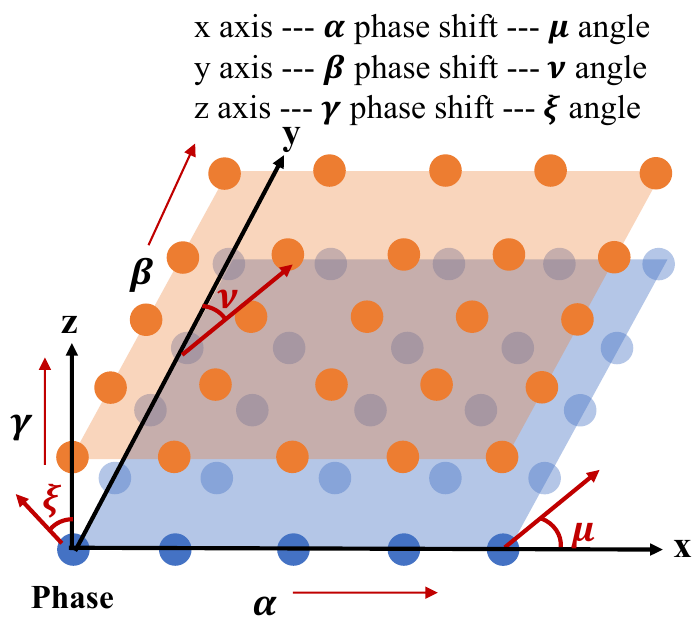}
    \caption{Geometry of the considered two layer 3D array.}
    \label{3DarrayGeometry}
    \vspace{-0ex}
\end{figure}

In this section, we focus on a two-layer 3D antenna array structure, where each layer is a 2D antenna array as analyzed in Section \uppercase\expandafter{\romannumeral2}. First, we analyze the case of infinite 2D arrays. Then, the discrete sampling method used to transition from infinite arrays to finite arrays can be similarly applied to analyze the performance limits of a finite 3D array.
For the 3D array structure, we denote its spatial radiation characteristics using a total of six parameters, divided into two groups: 
$(\alpha,\beta,\gamma)$ and $(\mu,\nu,\xi)$. Specifically, $(\alpha,\beta,\gamma)$ represent the linear phase changes between antenna elements along the 
$(x,y,z)$ axes, respectively. The corresponding angles between the in-phase directions and the $(x,y,z)$ axes are represented by 
$(\mu,\nu,\xi)$, as shown in Fig. \ref{3DarrayGeometry}. 

\par When the beam of the operating array antenna scans the space by varying the phases of element excitations, the presence of mutual coupling causes the signal reflected into each generator to vary. If all the coupling coefficients are available, the characteristics of the reflection signal versus the excitation phases can be computed. This is done simply by adding together the signals from all the generators as they are coupled back into one generator, as suggested in Fig. \ref{RC2Dinf} (a). Thus by reciprocity and superposition, the reflection coefficient of the considered 3D antenna array can be written as
\begin{equation}
    R(\alpha,\beta,\gamma) = \sum_{p=-\infty}^{+\infty}\sum_{q=-\infty}^{+\infty}\left[C_{pq}^{0}e^{j(p\alpha + q\beta)} + C_{pq}^{1}e^{j(p\alpha + q\beta + \gamma)}\right], 
    \label{active reflection coefficient 1}
\end{equation}
where the coupling terms $C_{pq}^{0}$ represent the mutual coupling coefficients between the reference antenna element and other antenna elements within the same 2D array plane, while $C_{pq}^{1}$ represent the coupling coefficients between the reference antenna element and the elements in the other planar layer. The relationship between the reflection coefficients and the coupling coefficients is illustrated in Fig. \ref{Determination of mutual coupling}. Due to the geometrical symmetry, we have $C_{pq}^{0} = C_{-p-q}^{0}$ and $C_{pq}^{1} = C_{-p-q}^{1}$ in infinitely large 3D array cases. Thus, each pair of elements that are symmetrical about the reference element contributes a term equal to $C_{pq}^{0}\exp\left[j\left(p\alpha + q\beta\right)\right] + C_{-p-q}^{0}\exp\left[j\left(-p\alpha - q\beta\right)\right] = \left(C_{pq}^{0}+ C_{-p-q}^{0}\right)\cos(p\alpha + q\beta) $. Similarly, the coupling between the reference antenna element and the elements of another planar antenna array follows the relationship 
\begin{equation}
    \begin{aligned}
        &C_{pq}^{1}\exp\left[j\left(p\alpha + q\beta + \gamma\right)\right] + C_{-p-q}^{1}\exp\left[j\left(-p\alpha - q\beta + \gamma\right)\right] \\
        & = \left(C_{pq}^{1}+ C_{-p-q}^{1}\right)\cos(p\alpha + q\beta) \exp(j\gamma),
    \end{aligned}
\end{equation}
by substituting the above symmetry relationships into the active reflection coefficient expression (\ref{active reflection coefficient 1}), we can obtain the following Fourier expansion
\begin{equation}
    R(\alpha,\beta,\gamma) = \sum_{p=-\infty}^{+\infty}\sum_{q=-\infty}^{+\infty}\left(C_{pq}^{0} + C_{pq}^{1}e^{j\gamma}\right)\cos\left(p\alpha + q\beta\right),
    \label{3D Fourier expansion relationship}
\end{equation}
and the corresponding Parseval's theorem can be derived as
\begin{equation}
    \begin{aligned}
        &\frac{1}{\pi^2}\iint_{0}^{\pi}  \left| R(\alpha,\beta,\gamma) \right|^2d\alpha d\beta = \sum_{p=-\infty}^{+\infty}\sum_{q=-\infty}^{+\infty}\left|C_{pq}^{0} + C_{pq}^{1}e^{j\gamma}\right|^2 \\
        & = \sum_{p=-\infty}^{+\infty}\sum_{q=-\infty}^{+\infty}\left(C_{pq}^{0} + C_{pq}^{1}e^{j\gamma}\right)\left(C_{pq}^{0*} + C_{pq}^{1*} e^{-j\gamma}\right) \\
        & = \sum_{p=-\infty}^{+\infty}\sum_{q=-\infty}^{+\infty}\left|C_{pq}^{0}\right|^2 + \left|C_{pq}^{1}\right|^2\\
        & \qquad \qquad \qquad + C_{pq}^{0}C_{pq}^{1*}e^{-j\gamma} + C_{pq}^{0*}C_{pq}^{1}e^{j\gamma}
    \end{aligned}
    \label{3D infnite energy relationship}
\end{equation}
where the first two terms are the squared magnitudes of the coupling coefficients needed to calculate the radiation efficiency limit of the 3D antenna array. The remaining two terms are the correlation of the coupling terms, which are not required. 
Through further observation, it is evident that when the inter-layer phase difference is set as $\gamma = 0$ and $\gamma = \pi$, the summation of the remaining two correlation terms can be written as 
$$ 
    \left\{  
        \begin{array}{ll}  
        C_{pq}^0C_{pq}^{1*} + C_{pq}^{0*}C_{pq}^1, & \gamma = 0, \\
        \\  
        -C_{pq}^0C_{pq}^{1*} - C_{pq}^{0*}C_{pq}^1,  & \gamma = \pi,
        \end{array}
    \right.  
$$
which can cancel each other out and be eliminated when summed. Therefore, by investigating the reflection coefficients of the 3D array under the conditions of $\gamma=0$ and $\gamma = \pi$, and then averaging the results, we can derive the expression for the radiation efficiency upper bound of the 3D antenna array. 
\begin{theorem}
    The embedded element efficiency for a two-layer infinite 3D array can be written as
    \begin{equation}
        \begin{aligned}
            \eta_{3D} &= 1-\sum_{p=-\infty}^{+\infty}\sum_{q=-\infty}^{+\infty} \left(\left|C_{pq}^0\right|^2 + \left|C_{pq}^1\right|^2\right) \\
            & = 1 - \frac{\frac{1}{\pi^2}\iint_{0}^{\pi}\left(\left|R(\alpha,\beta,0)\right|^2 + \left|R(\alpha,\beta,\pi)\right|^2\right)d\alpha d\beta }{2}
        \end{aligned}
        \label{3D efficiency infinite}
    \end{equation}
\end{theorem}
\begin{remark}
    The significance of Equation (\ref{3D efficiency infinite}) lies in indicating the average radiation efficiency limit of a 3D array when performing beamforming across the entire space. Specifically, this requires considering $\gamma$ as a parameter dependent on $(\alpha, \beta)$ to ensure that all antennas are in phase, and performing a double integral of Equation (\ref{3D Fourier expansion relationship}) with $(\alpha, \beta)$ as variables. However, analytically solving this integral is quite difficult. Therefore, in the above analysis, we treat $\gamma$ as a third free variable to obtain more intuitive theoretical analysis results. We have also performed numerical validations for (\ref{3D efficiency infinite}) and (\ref{3D efficiency finite}).
\end{remark}
\begin{remark}
    For a two-layer 3D array, we only need to evaluate the reflection coefficients for the two cases where the phase differences along the z-axis are 0 and 180 degrees. Then by taking the average, we can obtain the average coupling intensity and radiation efficiency of the 3D array. 
\end{remark}
\begin{remark}
    The physical interpretation of (\ref{3D efficiency infinite}) is that the energy coupled to other ports will be absorbed by the load and will not be radiated. 
\end{remark}

\par Furthermore, we can apply a similar derivation process as in Section \uppercase\expandafter{\romannumeral3}, where we transitioned from an infinite 2D antenna array to a finite 2D array. 
By discretely sampling the reflection coefficient plane $R(\alpha,\beta,\gamma)$, we can extend (\ref{3D efficiency infinite}) to a finite 3D antenna array scenario as follows:
\begin{equation}
    \begin{aligned}
        \eta_{3D}^{(MN)} &= 1-\sum_{p=0}^{M-1}\sum_{q=0}^{N-1} \left(\left|\overline{C}_{pq}^{0}\right|^2 + \left|\overline{C}_{pq}^{1}\right|^2\right) \\
        & =\frac{1}{2} - \frac{\sum_{\mu=0}^{M-1}\sum_{\nu=0}^{N-1}\left|R(\frac{2\pi\mu}{M},\frac{2\pi\nu}{N},0)\right|^2}{2MN} \\
        & + \frac{1}{2} - \frac{\sum_{\mu=0}^{M-1}\sum_{\nu=0}^{N-1} \left|R(\frac{2\pi\mu}{M},\frac{2\pi\nu}{N},\pi)\right|^2}{2MN}
    \end{aligned}
    \label{3D efficiency finite}
\end{equation}

\par Next, we conduct a more detailed analysis of the active reflection coefficient $R(\alpha,\beta,\gamma)$ of the considered 3D array structure. For infinitely large planar arrays, as analyzed earlier, there is no residual radiation in any out-of-phase condition since the major lobe occupies an infinitesimal angular region. 
Thus, infinitely large 3D arrays also exhibit radiation only in the main lobe direction. However, please note that the radiation beams from the two layers of the 3D array are not necessarily in-phase, depending on the setting of $\gamma$. Therefore, for the 3D array geometry shown in Fig. \ref{3DarrayGeometry}, the radiation direction is determined by the phase pair $(\alpha,\beta)$, but the radiation efficiency is determined by all three parameters $(\alpha,\beta,\gamma)$. Specifically, the phase delay induced by the transmission of EM waves with an angle $\theta$ to the $z$-axis is
\begin{equation}
    \delta_z = -2\pi\frac{d_z \cos\theta}{\lambda},
    \label{phase change 1}
\end{equation}
with a phase difference of $\gamma$ set between different layers, the phase difference between the beams from the two layers observed in the main lobe direction is
\begin{equation}
    \varphi = \gamma - 2\pi\frac{d_z \cos\theta}{\lambda},
    \label{phase change 2}
\end{equation}
and the corresponding active reflection coefficient can be written as
\begin{equation}
    \left|R(\alpha,\beta,\gamma)\right|^2 = 1 - \frac{\left|1+e^{j\varphi}\right|^2}{4} \in D
    \label{Reflection 3D}
\end{equation}
which corresponds to the percentage of energy that cannot be radiated, and $D$ is the 2D feasible region defined in (\ref{2D feasible region}).
The above analysis reveals the differences in spatial energy distribution between 3D and 2D arrays: for 2D arrays, energy is considered fully radiated if there is an in-phase radiation direction in space; otherwise, there is no radiation ($R(\alpha,\beta)$ only takes values of 0 or 1). However, the radiated energy varies with $\varphi$ for 3D arrays, and the corresponding $R(\alpha,\beta,\gamma)$ can take values in the range of $[0,1]$.
Based on (\ref{Reflection 3D}), the average reflection coefficient for a fixed $\gamma$ can be written as 
\begin{equation}
    \begin{aligned}
        & \hat{R}(\gamma) = \frac{1}{\pi^2}\iint_{0}^{\pi} \left|R(\alpha,\beta,\gamma)\right|^2 d\alpha d\beta\\
        & = 1 - \frac{1}{\pi^2}\iint_{D}\frac{1+\cos\varphi}{2} d\alpha d\beta\\
        & = 1 - \frac{\pi d_x d_y}{2\lambda^2}- \frac{1}{\pi^2}\iint_{D}\frac{\cos\left(\gamma -  \frac{ 2\pi d_z\cos\theta}{\lambda}\right)}{2}d\alpha d\beta,
    \end{aligned}
    \label{average R for fixed gamma}
\end{equation}
with $\cos\theta = \sqrt{1-\sin^2\theta} = \sqrt{1-\left(\frac{\lambda \alpha}{2\pi d_x}\right)^2 - \left(\frac{\lambda \beta}{2\pi d_y}\right)^2}$. Furthermore, by substituting (\ref{average R for fixed gamma}) into (\ref{3D efficiency infinite}), we have the following corollary:
\begin{corollary}
    The embedded element efficiency upper bound for a 3D array consisting of two parallel infinite planar uniform arrays can be formulated as
    \begin{equation}
        \begin{aligned}
            \eta_{3D} & = 1 - \frac{\frac{1}{\pi^2}\iint_{0}^{\pi}\left(\left|R(\alpha,\beta,0)\right|^2 + \left|R(\alpha,\beta,\pi)\right|^2\right)d\alpha d\beta }{2} \\
            & = 1- \frac{1 - \frac{\pi d_x d_y}{2\lambda^2}- \frac{1}{\pi^2}\iint_{D}\frac{\cos\left(0 -  \frac{ 2\pi d_z\cos\theta}{\lambda}\right)}{2}d\alpha d\beta}{2} \\
            & \quad\quad - \frac{1 - \frac{\pi d_x d_y}{2\lambda^2}- \frac{1}{\pi^2}\iint_{D}\frac{\cos\left(\pi -  \frac{ 2\pi d_z\cos\theta}{\lambda}\right)}{2}d\alpha d\beta}{2} \\
            & = \frac{\pi d_x d_y}{2\lambda^2} = \frac{1}{2} \eta_{2D},
        \end{aligned}
    \end{equation}
    \label{3D efficiency bound}
\end{corollary}

\vspace{-4ex}
\begin{remark}
    For infinitely large planar arrays, the embedded element efficiency of a 3D array decreases by half compared to an infinitely large 2D planar array due to the stronger coupling effects between the elements, as the 3D array has more antennas. 
\end{remark}
\begin{remark}
    Physical explanation: the gain of a 3D array is essentially derived from the increased effective projected area in the vertical dimension when observed from non-broadside directions. However, for infinitely large planar arrays, regardless of the finite spacing between the two layers, the increased projected area in the vertical dimension can always be neglected when observing the array from angles other than the broadside direction. Therefore, for the observer, the effective area of the antenna array is the same, and thus the corresponding array gain limit is also the same. However, since the number of antennas has doubled, the average radiation efficiency decreases by half accordingly.
\end{remark}
\begin{remark}
    Although there is no benefit in forming a 3D array from infinitely large planar arrays, a finite-size 3D array can achieve an increased effective aperture area in space compared to corresponding 2D arrays, thereby obtaining benefits. At this point, the efficiency performance of finite dimension 3D arrays can be evaluated using (\ref{3D efficiency finite}) or its approximation expression (\ref{3D Hannan approximation}) in the next subsection. Our subsequent simulations also demonstrate this point.
\end{remark}

\subsection{Gain Limit of Finite Aperture 3D Antennas}
\par As described in Section \uppercase\expandafter{\romannumeral2}, the gain limit of large 2D arrays is proportional to the projected area of the antenna aperture in the observation direction and can be approximated by  
$\frac{4\pi A_p}{\lambda^2}\cos\theta$ \cite{PHannan1964}.
For a 3D array, however, when observed from the spatial direction $(\theta,\phi)$, the expression for the total effective area is
\begin{equation}
    A_{e} = A_{xy}\cos\theta + A_{xz}\sin\phi\sin\theta + A_{yz}\cos\phi\sin\theta,
\end{equation}
therefore, for any spatial observation region $O = \left\{\theta_1 \leq \theta \leq \theta_2; \phi_1 \leq \phi \leq \phi_2\right\}$, the average gain limit of the 3D array normalized to a 2D array of the same aperture size is
\begin{equation}
    \resizebox{1\hsize}{!}{$
    \begin{aligned}
        &G = \frac{\iint_{D} A_{e,3D} \sin\theta d\theta d\phi}{\iint_{D} A_{e,2D}\sin\theta d\theta d\phi}, \\
        &= \frac{\iint_{D} \left(A_{xy}\cos\theta + A_{xz}\sin\phi\sin\theta + A_{yz}\cos\phi\sin\theta\right) \sin\theta d\theta d\phi }{ \iint_{D} A_{xy}\cos\theta \sin\theta d\theta d\phi}  \\
        & = 1+ \frac{A_{xz} \left(\cos \phi_1 -\cos \phi_2 \right) \left[ \theta_2 - \theta_1 - \frac{1}{2} \left( \sin 2\theta_2 - \sin 2\theta_1 \right) \right]}{A_{xy} \left(\phi_2 - \phi_1\right) \frac{1}{2} \left(\cos 2\theta_1 -\cos 2\theta_2 \right)} \\
        & \qquad + \frac{A_{yz} \left(\sin \phi_2 - \sin \phi_1\right) \left[ \theta_2 - \theta_1 - \frac{1}{2} \left( \sin 2\theta_2 - \sin 2\theta_1 \right) \right]}{A_{xy} \left(\phi_2 - \phi_1\right) \frac{1}{2} \left( \cos 2\theta_1 -\cos 2\theta_2 \right)},
    \end{aligned} $}
    \label{average gain limit}
\end{equation}
and the average gain of the 3D array over half space can be approximated by 
\begin{equation}
    G = 1 + \frac{A_{xz}+A_{yz}}{A_{xy}},
\end{equation}
thus, we can appropriate (\ref{3D efficiency finite}), i.e., the Hannan Limitation for finite-size 3D arrays, from the gain perspective as follows:
\begin{equation}
    \eta_{3D} =  \frac{N_{2D}}{N_{3D}}\left(1 + \frac{A_{xz}+A_{yz}}{A_{xy}}\right)\eta_{2D},
    \label{3D Hannan approximation}
\end{equation}
where $N_{3D}$ and $N_{2D}$ represent the total number of antennas of finite-size 3D and 2D arrays, respectively.
\par For practical base station antenna arrays, the horizontal hemispace scanning gain is often more important than the vertical hemispace scanning gain. This is because, in real communication scenarios, users are usually randomly distributed in the horizontal plane. However, in the vertical plane, apart from special scenarios such as base stations near high-rise buildings, the demand for wide-range vertical beamforming is rare. Therefore, considering horizontal hemispace scanning $(\theta_1 = 0, \theta_2 = \frac{\pi}{2}; \phi_1 = \phi_2 = \frac{\pi}{2})$, the average gain of the 3D array is
\begin{equation}
    G = 1 + \frac{A_{xz}}{A_{xy}},
\end{equation}
please note that the above formula is an approximate analysis of the upper limit of the achievable gain for 3D antenna arrays. In this analysis, we consider the projected area of the 3D antenna array in the observation direction as its effective aperture area. In the subsequent simulation section, we validate the effectiveness of this approximation method by examining the hemispherical average gain of actual 3D antenna arrays. However, a rigorous derivation of the effective aperture should be related to the maximum achievable directivity of the array \cite[Eqn. (2-110)]{Balanis2016}.
\begin{equation}
    A_{em} = \frac{\lambda^2}{4\pi}D_0,
\end{equation}
where $A_{em}$ and $D_0$ represent the maximum effective aperture and the corresponding maximum directivity of any antenna, respectively.

\section{3D Feasible Region}
\par In this section, we will analyze the feasible region of the $(\alpha,\beta)$ plane for the proposed two-layer 3D antenna array. The 3D array structure we consider is shown in Fig. \ref{3DarrayGeometry}. 
As analyzed in (\ref{phase change 1}), we assume that $(\alpha,\beta)$ produces a spatial beam direction with an angle $\theta$ to the $z$-axis in the first octant of space, and the phase change induced by this beam as it propagates from the lower planar array to the upper planar array is given by $\delta_z = -2\pi\frac{d_z \cos\theta}{\lambda}$. 
Meanwhile, the phase difference $\gamma$ set between adjacent elements along the $z$-axis in the 3D array may not match $\delta_z$, leading to a reduction in array gain and spatial radiation energy. Here, we rewrite the phase difference between the upper and lower arrays caused by this phase mismatch as
\begin{equation}
    \varphi = \frac{2\pi d_z}{\lambda}\left(\cos\xi - \cos\theta\right), 
    \label{phiexpression}
\end{equation}
where $\xi$ represents the angle between $z$ axis and the add-in phase direction determined by $\gamma$. 
\par
\begin{figure}[!ht]
    \vspace{-2ex}
    \centering
    \includegraphics[width=3in]{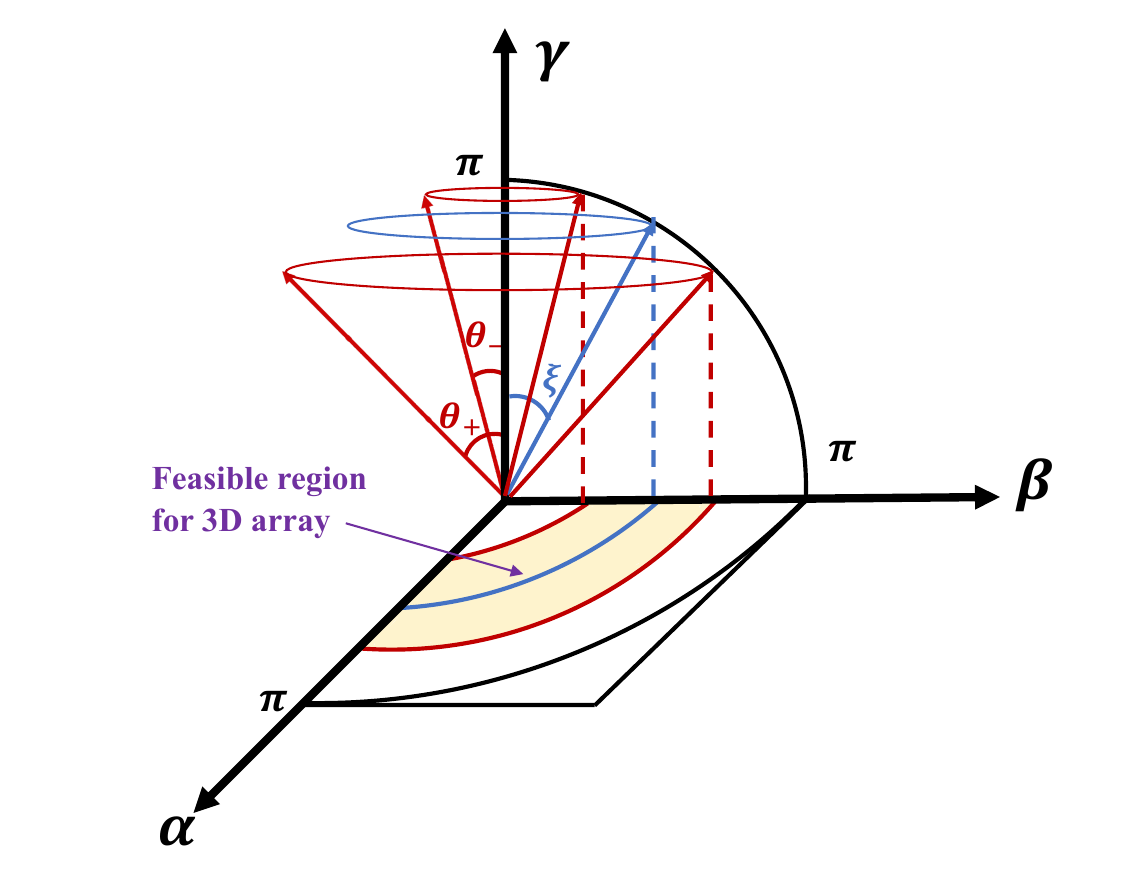}
    \caption{Diagram of the planar feasible region and the beamforming direction relationship for 3D arrays.}
    \label{3D feasible Region}
    \vspace{0ex}
\end{figure}

 Assume that the energy radiated by a single array element is considered as one unit. When the phase between the upper and lower array planes is perfectly matched (i.e., $\xi = \theta$), the radiated electric field intensity is two, and the energy intensity is four times that radiated by a single 2D array. In the presence of a phase mismatch $\varphi$, the total radiated electric field intensity is $\left| 1 + e^{j\varphi} \right|$, and the energy becomes $\left| 1 + e^{j\varphi} \right|^2 = 2 + 2\cos\varphi$ times that of a single 2D array. Please note that our theory can be easily extended from a two-layer 3D antenna array to a multi-layer 3D array scenario. When the 3D array consists of $L_z$ layers, the expression $\left| 1 + e^{j\varphi} \right|$ needs to be modified to $\left| \sum_{l=1}^{L_z} e^{jl\varphi} \right|$, with the subsequent analysis remaining the same. From the energy perspective, we introduce an attenuation threshold $t$ such that $ t = \left| 1+e^{j\varphi_t} \right| = \sqrt{2+2\cos\varphi_t}$. When the electric field intensity exceeds $t$, it is considered a feasible beam; otherwise, it is deemed an infeasible beam. From the definition of the energy threshold $t$, it can be derived that
\begin{equation}
    \cos\varphi_t = \frac{t^2-2}{2},
    \label{cosphiexpression}
\end{equation}
by combining (\ref{phiexpression}) and (\ref{cosphiexpression}), we can derive that under the condition of a fixed phase shift along the $z$-axis, the boundary of $\theta$ that satisfies the amplitude threshold $t$ is given by
\begin{equation}
    \cos\left[  \frac{2\pi d_z}{\lambda}\left(\cos\xi - \cos\theta\right) \right] = \frac{t^2-2}{2},
\end{equation}
which can be further decomposed into two boundary conditions for $\theta$
\begin{equation}
    \begin{aligned}
        &\cos\theta_{-} - \cos\xi = \frac{\lambda}{2\pi d_z}\arccos\left(\frac{t^2-2}{2}\right) \\
        &\cos\xi - \cos\theta_{+} = \frac{\lambda}{2\pi d_z}\arccos\left(\frac{t^2-2}{2}\right),
    \end{aligned}
\end{equation}
and the corresponding feasible $\theta$ region can be formulated as
\begin{equation}
    \begin{aligned}
        &\theta \geq \theta_{-} = \arccos\left[\cos\xi+\frac{\lambda}{2\pi d_z}\arccos\left(\frac{t^2-2}{2}\right)\right] \\
        &\theta \leq \theta_{+} = \arccos\left[\cos\xi-\frac{\lambda}{2\pi d_z}\arccos\left(\frac{t^2-2}{2}\right)\right],
        \label{feasible theta}
    \end{aligned}
\end{equation}
which is shown in Fig. \ref{3D feasible Region}. 

\par
\begin{figure}[!ht]
    \vspace{0ex}
    \centering
    \includegraphics[width=2.6in]{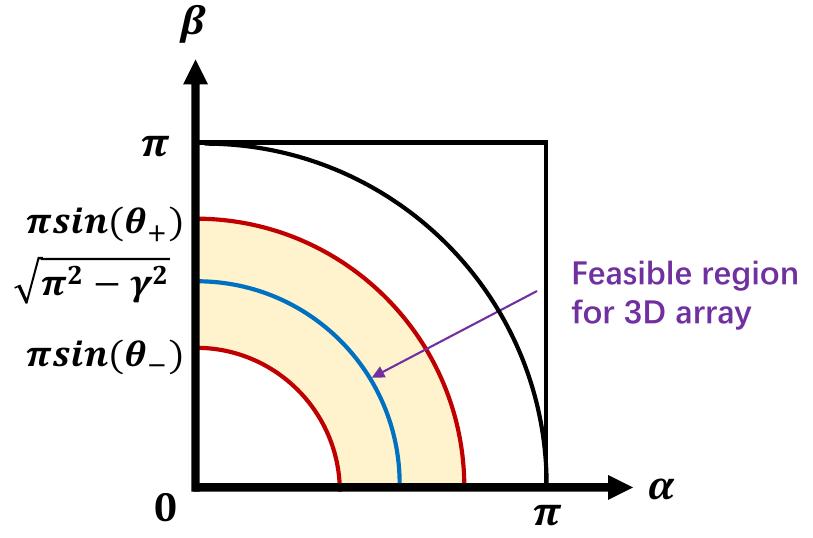}
    \caption{Planar feasible region for fixed amplitude parameter $t$ with  $d_x = d_y = d_z = \frac{\lambda}{2}$.}
    \label{planarfeasibleregion1}
    \vspace{-0ex}
\end{figure}

 From the derivation of the 2D Hannan Limitation, we know that the beamforming direction of a 2D array and the conical angles $(\mu,\nu)$ with respect to the $x$ and $y$ coordinate axes satisfy the relationship $\sin^2\theta = \cos^2\mu + \cos^2\nu$. Therefore, the feasible region for the planar phase set $(\alpha, \beta)$ is given by
\begin{equation}
    \left\{
            \begin{array}{lr}
                \cos^2(\mu) + \cos^2(\nu) \geq \sin^2(\theta_{-}), \\
                \cos^2(\mu) + \cos^2(\nu) \leq \sin^2(\theta_{+}),
            \end{array}
    \right. 
\end{equation}
which can be expanded as 
\begin{equation}
    \resizebox{0.89\hsize}{!}{$
    \left\{
            \begin{array}{lr}
                \left(\frac{\alpha\lambda}{2\pi d_x}\right)^2 + \left(\frac{\beta\lambda}{2\pi d_y}\right)^2\geq 1- \left[\cos\xi+\frac{\lambda}{2\pi d_z}\arccos\left(\frac{t^2-2}{2}\right)\right]^2, \\
                \left(\frac{\alpha\lambda}{2\pi d_x}\right)^2 + \left(\frac{\beta\lambda}{2\pi d_y}\right)^2 \leq  1- \left[\cos\xi-\frac{\lambda}{2\pi d_z}\arccos\left(\frac{t^2-2}{2}\right)\right]^2,
            \end{array}
    \right. 
    $}
    \label{feasiblering}
\end{equation}
the above expansion is due to $\cos\mu = \frac{\alpha\lambda}{2\pi d_x}$ and $\cos\nu = \frac{\beta\lambda}{2\pi d_y}$. 
From (\ref{feasiblering}), we can see that the general feasible region for a fixed amplitude parameter $t$ forms a ring shape, as illustrated in Fig. \ref{planarfeasibleregion1}, where we set $d_x = d_y = d_z =\frac{\lambda}{2}$.

\par In the following, we will analyze UPA arrays with $d = d_x = d_y \leq \frac{\lambda}{2}$. In this case, the feasible region boundary simplifies from a more general elliptical shape to a more easily analyzable circular shape.
Based on the above discussions, for any given phase shift parameter $\gamma$  and amplitude attenuation parameter $t$, we can derive the analytical expressions for the area of the feasible region in the $(\alpha, \beta)$ plane, as well as the inner and outer radii of this area.
\begin{theorem}
    The feasible region of a 2-layer infinite 3D array for a fixed $\xi$ is 
    \begin{equation}
        \resizebox{0.89\hsize}{!}{$
        \begin{aligned}
            &r_{-} = \sqrt{\left(\frac{2\pi d}{\lambda}\right)^2 - \left[ \min\left(\frac{2\pi d}{\lambda}\cos\xi + \arccos\left(\frac{t^2-2}{2}\right),\frac{2\pi d}{\lambda}\right) \right]^2}\\
            &r_{+} = \sqrt{\left(\frac{2\pi d}{\lambda}\right)^2 - \left[ \max\left(\frac{2\pi d}{\lambda}\cos\xi - \arccos\left(\frac{t^2-2}{2}\right),0\right) \right]^2}\\
            &S(\gamma,t) = \frac{2\pi^2 d}{\lambda}\cos\xi\arccos\left(\frac{t^2-2}{2}\right),
        \end{aligned}
        $}
    \end{equation}
\end{theorem}
\noindent From the above theorem, we can derive the normalized 3D feasible volume by integrating  $\gamma$ with the corresponding feasible region area.
\begin{equation}
    \begin{aligned}
        V(t) &= \frac{1}{\pi^3}\int_{0}^{\pi} S(\gamma,t)d\gamma
    \end{aligned}
    \label{feasible volumn}
\end{equation}

\par However, although we have theoretically demonstrated that the feasible plane region where a 3D array can generate a relatively concentrated radiation beam follows a ring shape, it is not the case that for all values of $\gamma$, the upper and lower bounds of the feasible range are simultaneously within the first quadrant. Specifically, for certain settings of $\gamma$ that make $\xi$ close to 0 degrees or 90 degrees (corresponding to $\gamma = \frac{2\pi d_z}{\lambda}$ and $\gamma = 0$, respectively), $\theta_{-}$ and $\theta_{+}$ do not both fall within the range of 0 to 90 degrees. This requires further detailed investigation. Specifically, we analyze the joint feasible range of  $(\alpha,\beta)$ in the following three cases.

\begin{figure}[!ht]
    \vspace{0ex}
    \centering
    \includegraphics[width=2.6in]{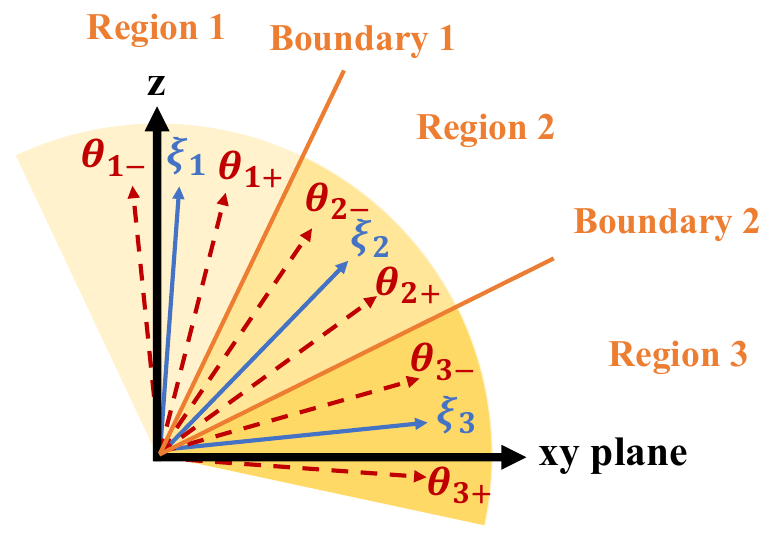}
    \caption{Different planar feasible region cases dependent on $\xi$.}
    \label{FeasibleRegionCase}
    \vspace{-1ex}
\end{figure}

\subsection{Around z axis}
\par When the beamforming angle $\xi$ is small and around the $z$-axis, the corresponding left boundary $\theta_{-}$ may fall within the second quadrant. In this case, the relationship $\cos\theta_{-} = \cos\xi + \frac{\lambda}{2\pi d_z}\arccos(\frac{t^2-2}{2})$ no longer holds. For example, when $\xi_{-} = 0$, for large threshold values $t$, the result of the above equation may be greater than 1, which can not represent the physical boundary between the feasible range and the $z$-axis. However, we observe that since $\xi \geq 0$ and the feasible range exhibits widening in both directions with respect to the beamforming direction specified by $\xi$, the right boundary $\theta_{+}$ must be greater than the left boundary $\theta_{-}$. Thus, the planar feasible range defined by $\theta_{+}$ is guaranteed to include the planar feasible range corresponding to $\theta_{-}$. Considering that we only consider the cases where $\alpha$ and $\beta$ are within the range $[0,\pi]$, the final planar feasible range in this situation is constrained by the x-axis, y-axis, and $\frac{\left(\alpha^2+\beta^2\right)\lambda^2}{\left(2\pi d\right)^2} \leq \sin^2\theta_{+}$. It forms a quarter-circle within the first quadrant, as shown in region 1 of Fig. \ref{FeasibleRegionCase} and Fig. \ref{Feasible region analysis} (b).

\subsection{Medium case}
\par We have already discussed this typical situation in our previous analysis of the planar feasible range for a 3D antenna array and its comparison with the feasible range of planar arrays, as shown in region 2 of Fig. \ref{FeasibleRegionCase}. Please note that in this case, the feasible angular width of the planar beamforming, i.e., $\theta_{+}-\theta_{-}$, is determined by $t$. Lareger values of $t$ result in narrower feasible angular widths.

\subsection{Around xy plane}
\par In contrast to the case discussed in subsection \textit{A}, we will now consider the situation that the beamforming angle $\xi$ is close to 90 degrees  and lies around the $xy$ plane (corresponding to a near-zero phase shift $\gamma$ along the $z$-axis). In this scenario, the corresponding right boundary $\theta_{+}$ may fall within the fourth quadrant. Please note that although the relationship $\cos\theta_{+} = \cos\xi - \frac{1}{\pi}\arccos(\frac{t^2-2}{2})$ still holds, we cannot consider the integration domain in the same way as in the second-class classical case where we set $\frac{\left(\alpha^2+\beta^2\right)\lambda^2}{\left(2\pi d\right)^2} \leq \sin^2\theta_{+}$ as an upper bound. 
To determine the range of the planar feasible region in this case, we further divide it into two scenarios: $\frac{\left(\alpha^2+\beta^2\right)\lambda^2}{\left(2\pi d\right)^2} \leq 1$ and $\frac{\left(\alpha^2+\beta^2\right)\lambda^2}{\left(2\pi d\right)^2} > 1$.


\par Firstly, the region satisfying $ \sin^2\theta_{-} \leq\frac{\left(\alpha^2+\beta^2\right)\lambda^2}{\left(2\pi d\right)^2} \leq 1$ is feasible. The underlying physical phenomenon is that the beamforming direction formed by the $(\alpha,\beta)$ phase set lies within the first octant of the 3D space. However, as shown in region 3 of Fig. \ref{FeasibleRegionCase}, when $\theta$ is between 90 degrees and $\theta_{+}$, the corresponding physical phenomenon is that the beamforming direction lies outside the first octant. This indicates that at the 90-degree boundary, there is still a residual energy attenuation margin. Therefore, some points in the region where $\frac{\left(\alpha^2+\beta^2\right)\lambda^2}{\left(2\pi d\right)^2}$ is greater than 1 are also feasible. The area of this region can be estimated  as the plane feasible region corresponding to $\sin^2\theta_{+} \leq \frac{\left(\alpha^2+\beta^2\right)\lambda^2}{\left(2\pi d\right)^2} \leq 1$ based on the symmetry, but its spatial radiation direction cannot be accurately controlled, thus it has limited practical application significance.

\subsection{Number and Boundaries of Feasible Regions}
\par Based on the above analysis, we can further derive a set of 3D phase settings that cover the entire space for a given threshold. Specifically, as shown in Fig. \ref{FeasibleRegionCase}, we divide the 3D space into different regions, each connected end to end. When we need to radiate energy into the 3D space, we can select the corresponding $\xi$ parameter based on the elevation angle and the corresponding $(\alpha,\beta)$ parameters based on the azimuth angle. Since the main new feature of the 3D array is its $\xi$ parameter, we focus on further research and design for this parameter.

\par Assuming that under a given attenuation threshold $t$, we need to cover the entire space with $P$ discrete \(\xi\). Specifically, we determine the first codeword \(\xi_{1}\) by setting \(\theta_{1-}=0\). Subsequently, we determine \(\xi_{2}, \ldots, \xi_{P}\) sequentially by setting \(\theta_{p-1,+} = \theta_{p,-}\) until \(\theta_{P,+} > \frac{\pi}{2}\). From (\ref{feasible theta}), we have
\begin{equation}
    \begin{aligned}
        \theta_{p-1,+} &= \arccos\left[\cos\xi_{p-1}-\frac{\lambda}{2\pi d_z}\arccos\left(\frac{t^2-2}{2}\right)\right] \\
        \theta_{p,-} &= \arccos\left[\cos\xi_{p}+\frac{\lambda}{2\pi d_z}\arccos\left(\frac{t^2-2}{2}\right)\right]
    \end{aligned}
\end{equation}
by setting \(\theta_{p-1,+} = \theta_{p,-}\), we have $\cos\xi_{p} = 
\cos\xi_{p-1}-\frac{\lambda}{\pi d_z}\arccos(\frac{t^2-2}{2})$. 
Moreover, since the first $\xi$ can be determined by $\cos\xi_{1} = \cos\theta_{1-} - \frac{\lambda}{2\pi d_z}\arccos(\frac{t^2-2}{2}) = 1 - \frac{\lambda}{2\pi d_z}\arccos(\frac{t^2-2}{2})$, we have 
\begin{equation}
    \cos\xi_{p} = 1 - \frac{(2p-1)\lambda}{2\pi d_z}\arccos\left(\frac{t^2-2}{2}\right),
\end{equation} 

\par The final spatial region $P$ can be classified into two scenarios: (1) $\left\{\xi_{P} \leq \frac{\pi}{2}, \theta_{P,+} \geq \frac{\pi}{2}\right\}$; (2) $\left\{\theta_{P-1,+} \leq \frac{\pi}{2}; ,\xi_{P} \geq \frac{\pi}{2} \right\}$. And the corresponding ranges for $P$ in the above two scenarios are 
\begin{equation}
    \left\{
            \begin{array}{lr}
                2P-1 \leq \frac{2\pi d_z}{\lambda \arccos\left(\frac{t^2-2}{2}\right)} \leq 2P, \vspace{1ex} \\ 
                2P-2 \leq \frac{2\pi d_z}{\lambda \arccos\left(\frac{t^2-2}{2}\right)} \leq 2P-1,
            \end{array}
    \right. 
\end{equation}
and the total number of feasible regions $P$ can be written as
\begin{equation}
    P = \lfloor \frac{2\pi d_z}{2 \lambda \arccos(\frac{t^2-2}{2})} \rfloor + 1,
\end{equation}

\section{Numerical Results}

\par
\begin{figure}[!ht]
    \vspace{0ex}
    \centering
    \includegraphics[width=3in]{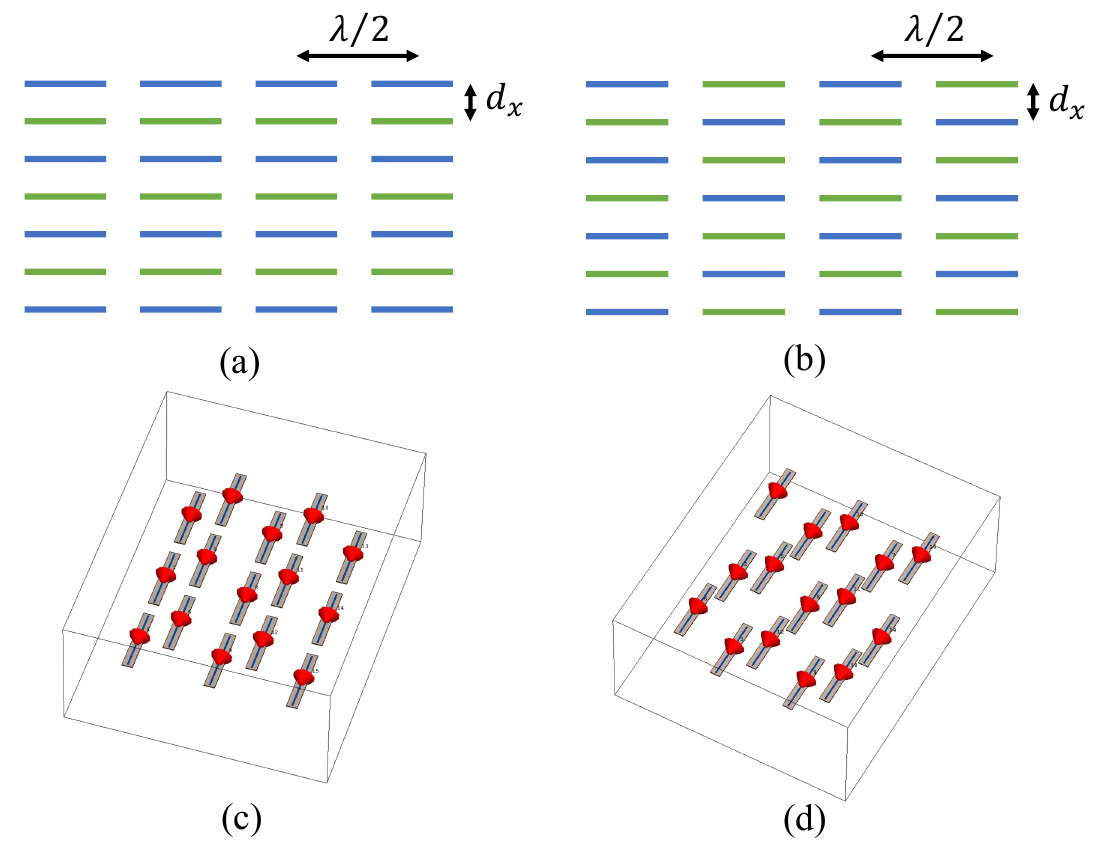}
    \caption{3D Array Geometry.}
    \label{3D Array Geometry}
    \vspace{-0ex}
\end{figure}

 In this section, simulation results are presented to validate our theories and evaluate the performances of the proposed 3D array. 
The specific configurations are as follows: the communication system is assumed to operate at 1.6 GHz. The array is composed of identical printed dipole antennas, which are printed on an FR-4 substrate measuring $12.2 \enspace \text{mm} \times 78 \enspace \text{mm}$, and the dimension of each dipole element is $1 \enspace \text{mm} \times 71.5 \enspace \text{mm}$.
Regarding the 3D array topology, to reduce mutual coupling between antenna elements, we adopt two horizontally staggered 3D antenna structures, as shown in Fig. \ref{3D Array Geometry}.

\par The dipole antennas are arranged along the $y$-axis at a fixed half-wavelength spacing, but the spacing along the $x$-axis is variable. For array simulations with a fixed array aperture area of $L_x\times L_y$, we maintain the antenna spacing at $0.5\lambda$ along the $y$-axis and set the antenna spacing along the $x$-axis as $\frac{L_x}{M-1}$, where $M$ is the number of antennas along the $x$-axis. For array simulations with fixed antenna spacing settings, unless otherwise specified, the default setup is an $M \times M$ square setting for the lower-layer 2D antenna array, with an antenna spacing of $0.5\lambda$ along both the $x$-axis and $y$-axis.


\begin{figure}[!ht]
    \vspace{0ex}
    \centering
    \includegraphics[width=3in]{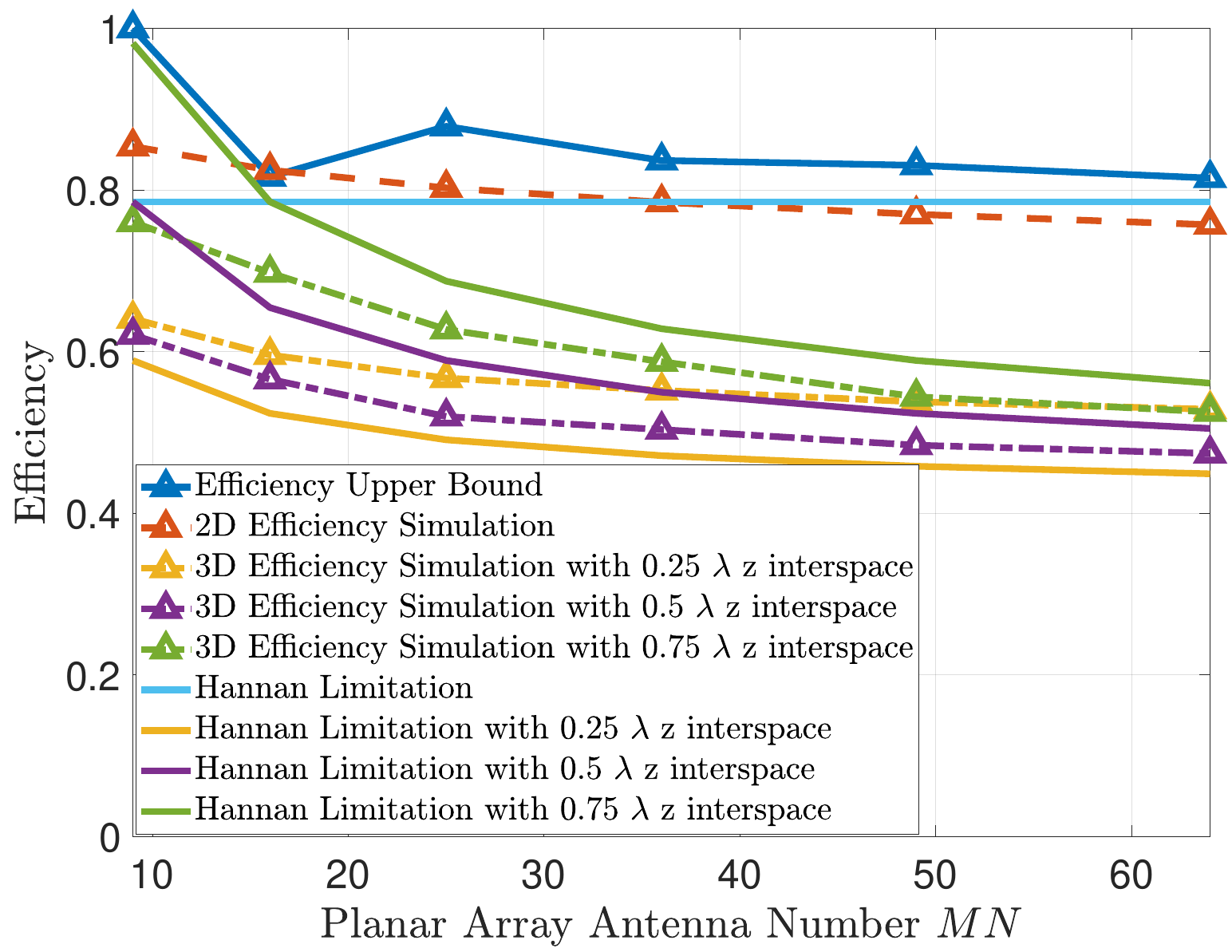}
    \caption{Radiation Efficiency for 2D/3D Array with $0.5\lambda$ spacing.}
    \label{Temp1}
    \vspace{-0ex}
\end{figure}
\par Fig. \ref{Temp1} displays the average radiation efficiency for 2D and 3D arrays with different inter-layer spacing configurations. It can be observed that Hannan Limitation is well applicable for large 2D arrays, which aligns with the simulation and measurement results reported in \cite{Kildal2015, SYuan2023}. Additionally, the proposed theoretical radiation efficiency limit for finite-dimensional 2D arrays consistently remains above that of the actual 2D arrays. It also decreases with increasing array size and ultimately converges to the Hannan Limitation.
In terms of 3D arrays, due to the inclusion of a larger number of antenna elements and the corresponding coupling effects, the overall efficiency of 3D arrays is lower than that of equivalent aperture 2D arrays. Moreover, the proposed 3D Hannan Limitation approximation formula (\ref{3D Hannan approximation})  provides a tight efficiency upper bound on radiation for large 3D arrays. The inaccuracy of the proposed estimation formula at a $0.25\lambda$ spacing is due to the strong distortions on element radiation patterns due to close inter-layer spacing, making the $\cos\theta$ pattern assumption in \cite{PHannan1964} no longer valid.

\begin{figure}[!ht]
    \vspace{0ex}
    \centering
    \includegraphics[width=3in]{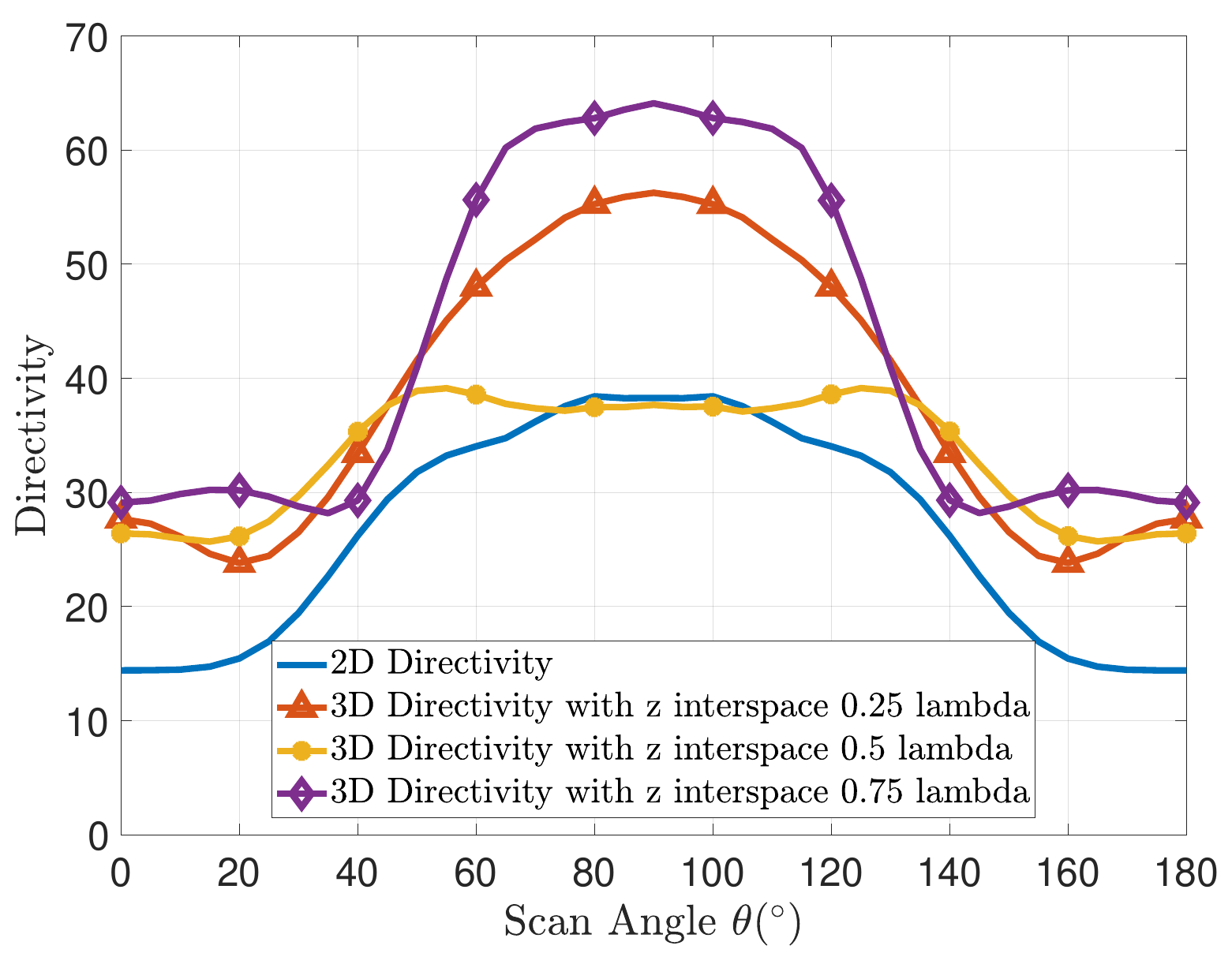}
    \caption{Directivity performances of the 2-D and 3-D arrays.}
    \label{directivity_vs_halfspacescan}
    \vspace{-0ex}
\end{figure}

\par Fig. \ref{directivity_vs_halfspacescan} shows the achieved directivity of 2D and 3D arrays with different interlayer spacings for various azimuth scan angles in the $xz$-plane. It can be observed that from a beamforming perspective, the 3D array has better performance than the 2D case, especially at large scanning angles, due to the larger projection area of the array. Furthermore, it can be seen that the array directivity at large scanning angles is proportional to the interlayer spacing in the 3D array. Finally, when the interlayer spacing of the 3D array is $0.5\lambda$, the directivity of the 3D array equals that of the 2D array near a scan angle of $90^{\circ}$ because both arrays produce double beams. However, the 3D array with interlayer spacings of $0.25\lambda$ and $0.75\lambda$ can produce a single beam at a scan angle of $90^{\circ}$.

\begin{figure}[!ht]
    \vspace{0ex}
    \centering
    \includegraphics[width=3in]{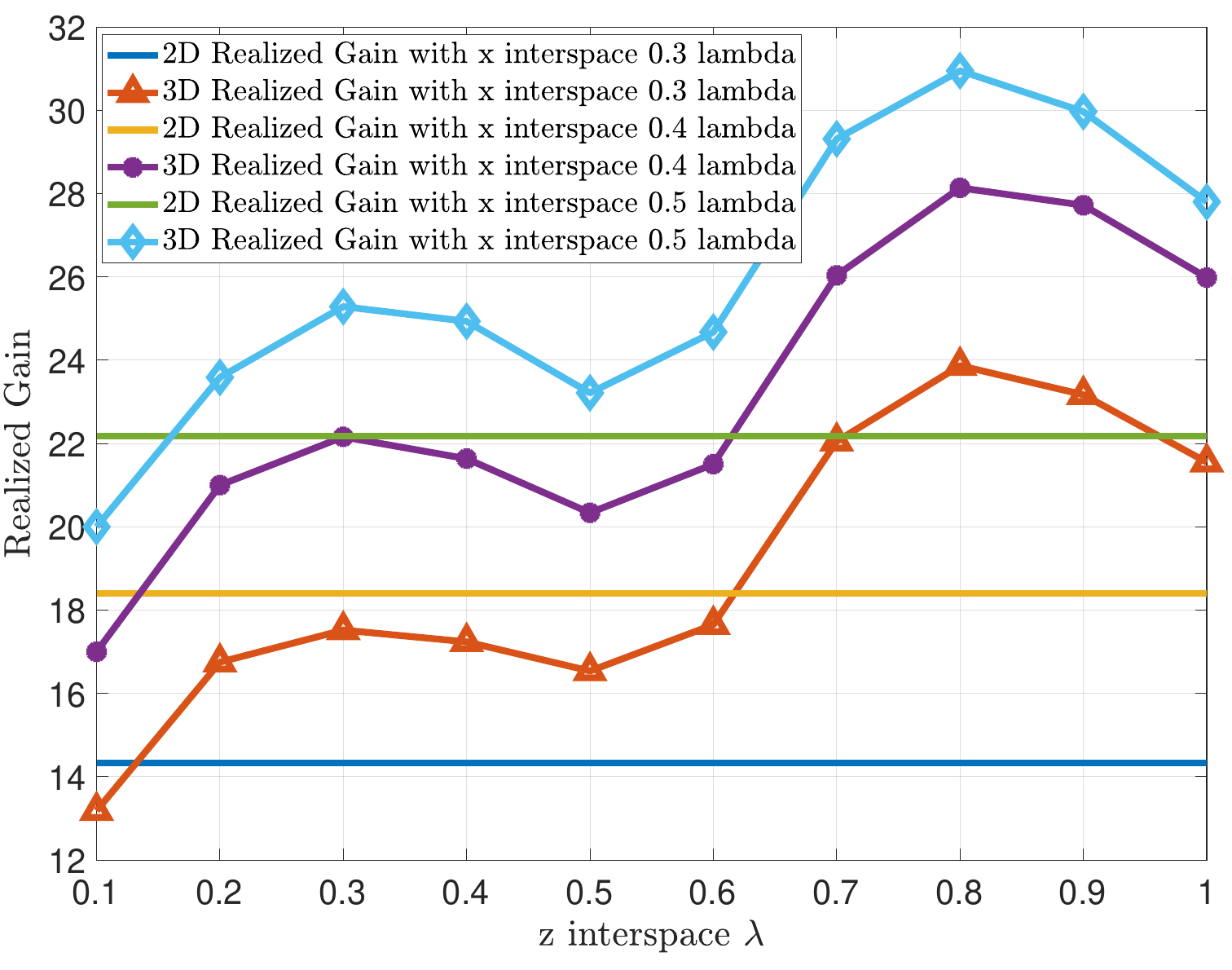}
    \caption{Average realized gain of 3D array over half-space with different inter-layer spacings.}
    \label{Temp2}
    \vspace{-2ex}
\end{figure}
\par Fig. \ref{Temp2} shows the variation of the hemispace average beamforming gain of 3D arrays as the inter-layer spacing increases. 
It can be seen that the antenna arrays exhibit local minimum points in average realized gain at spacings of $0.1\lambda$, $0.5\lambda$, and $1\lambda$. The first point is due to the strong coupling effect and the rest two points reflect the impact of the coherent energy cancellation effect introduced by the reflection effect between the two-layer planar arrays (due to the $\pi$ phase shift introduced by PEC reflection, the realized gain decreases for inter-layer spacings that are multiples of $0.5\lambda$, and the proportion of reflected energy increases). Therefore, there exists an optimal geometric structure for 3D antenna arrays, and it should be designed considering the electromagnetic effects.

\begin{figure}[!ht]
    \vspace{0ex}
    \centering
    \includegraphics[width=3in]{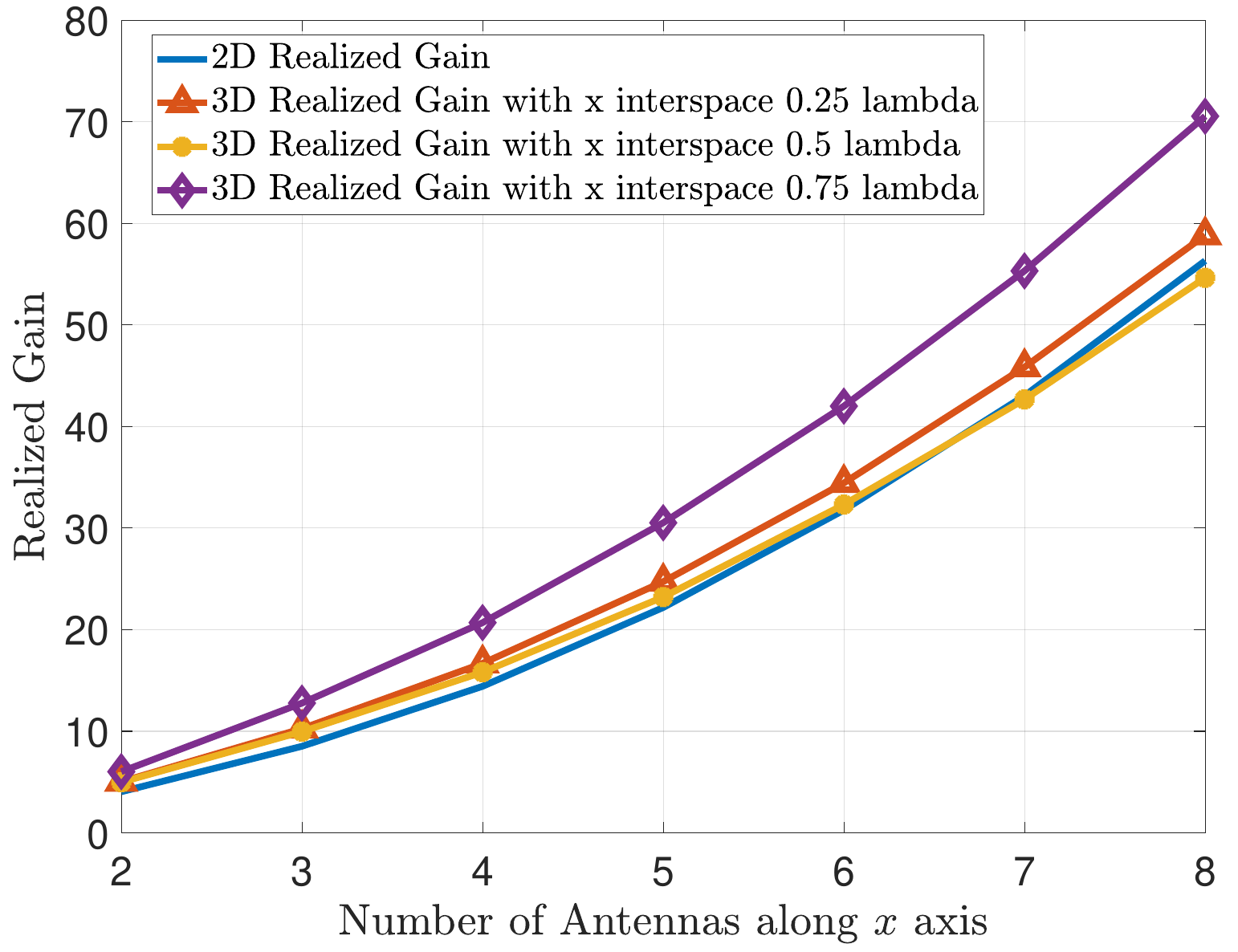}
    \caption{Average realized gain of 3D array over half-space with fixed $0.5\lambda$ interspacing.}
    \label{Temp3}
    \vspace{-3ex}
\end{figure}
\par Fig. \ref{Temp3} illustrates how the average realized gain in the hemispace of a 3D array with $M = N$ varies with the number of antennas, under the condition that the antenna spacing is fixed at $0.5\lambda$ (where the aperture area of the array increases according to a square law). It can be observed that the realized gain of both 2D and 3D arrays increases with the number of antenna elements. However, a significant performance gain only occurs when the inter-layer spacing of the 3D antenna array is increased to $0.75\lambda$.

\begin{figure}[!ht]
    \vspace{0ex}
    \centering
    \includegraphics[width=3in]{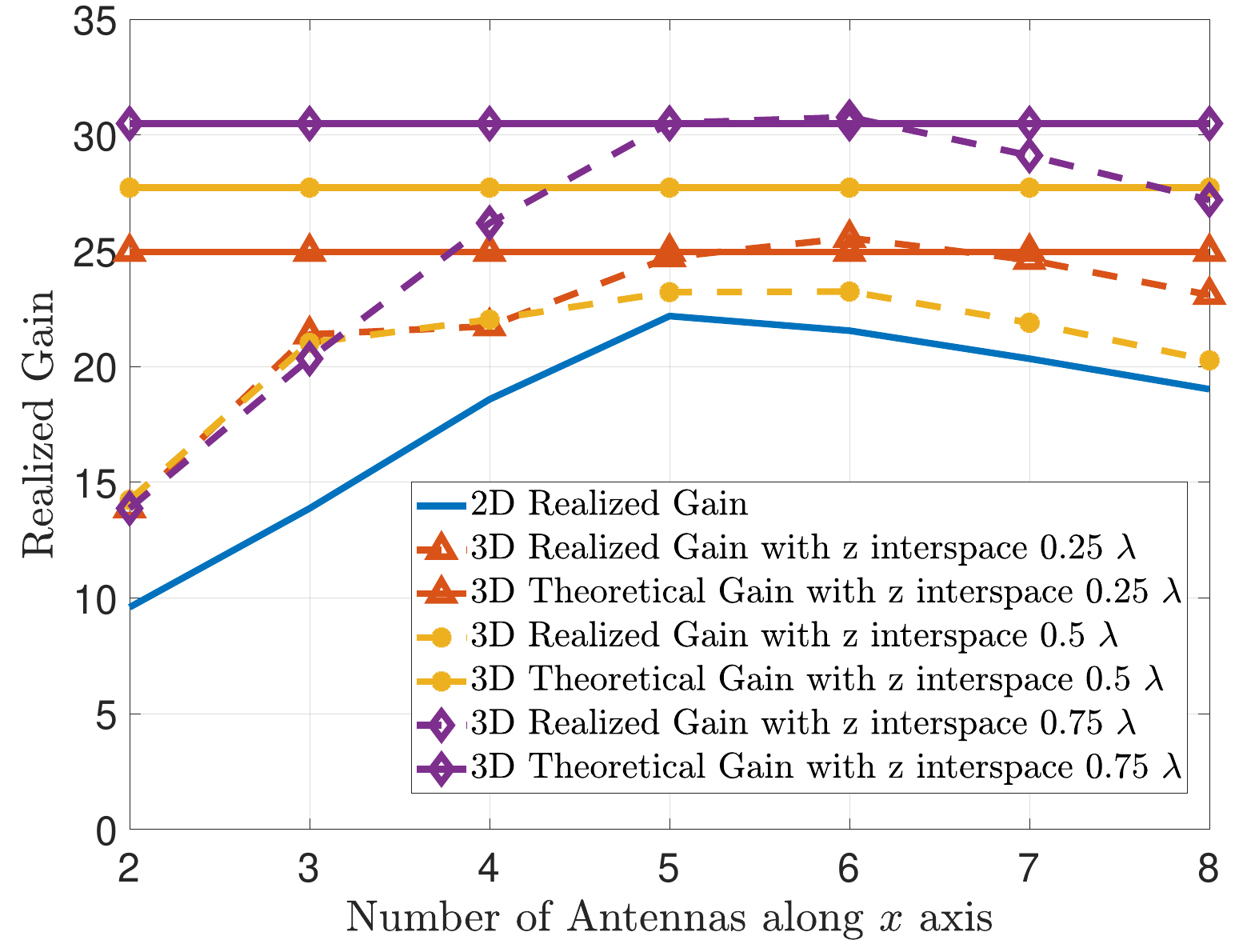}
    \caption{Average realized gain of 3D array (Fig. \ref{3D Array Geometry} (a) geometry) over half-space with fixed aperture area.}
    \label{Temp4}
    \vspace{-1ex}
\end{figure}

\par As a contrast to Fig. \ref{Temp3}, Fig. \ref{Temp4} shows the average realized gain for different numbers of antennas with a fixed aperture area of $2\lambda \times 2\lambda$. It is observed that the 3D array exhibits a more pronounced performance gain compared to the 2D array. Additionally, when the array antenna arrangement becomes overly dense, the realized gain decreases as EM mutual coupling effects becoming increasingly evident, resulting in a lower radiation efficiency. Therefore, for the two-layer 3D array considered in this paper, there also exists an optimal choice for the horizontal spacing between antenna elements. Moreover, if we do not consider a linear phase shift array and leverage EM coupling effects, a smaller optimal antenna spacing can be achieved \cite{Ranji2024,LHanSuperdirectiveMultiuser}.

\begin{figure}[!ht]
    \vspace{0ex}
    \centering
    \includegraphics[width=3in]{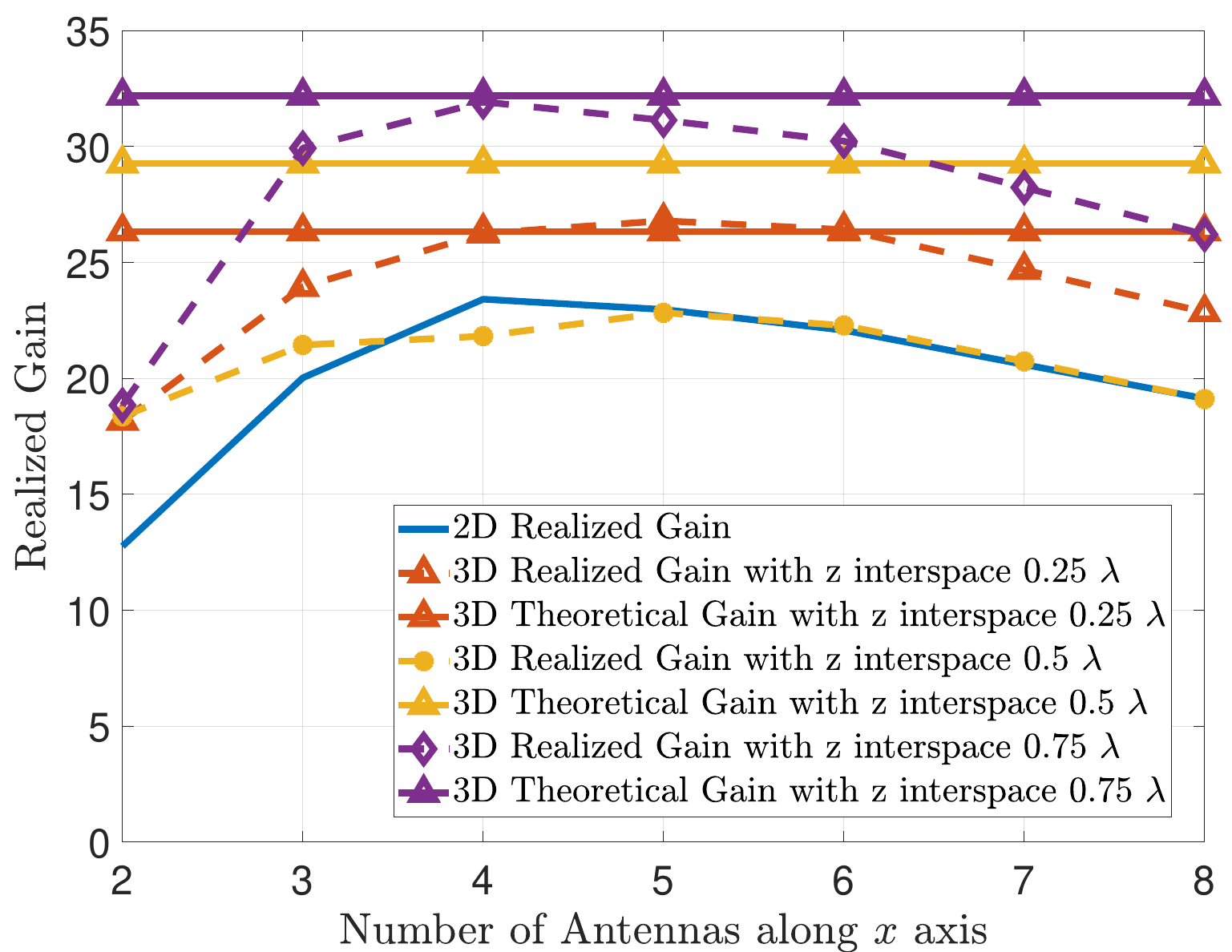}
    \caption{Average realized gain of 3D array (Fig. \ref{3D Array Geometry} (b) geometry) over half-space with fixed aperture area.}
    \label{Temp5}
    \vspace{-3ex}
\end{figure}
\par Fig. \ref{Temp5} shows the average beamforming gain in the hemispace for the 3D antenna array structure presented in Fig. \ref{3D Array Geometry} (b) while maintaining $2\lambda \times 2\lambda$ aperture area. Compared to Fig. \ref{Temp4}, it can be seen that approximately the same maximum achievable gain is realized with different 3D array arrangements. This is because the projection area of these two structures is the same. However, due to the larger spacing between antennas in the arrangement shown in Fig. \ref{3D Array Geometry} (b), the maximum gain can be achieved with fewer antenna elements.

\begin{figure}[!ht]
    \vspace{0ex}
    \centering
    \includegraphics[width=3.2in]{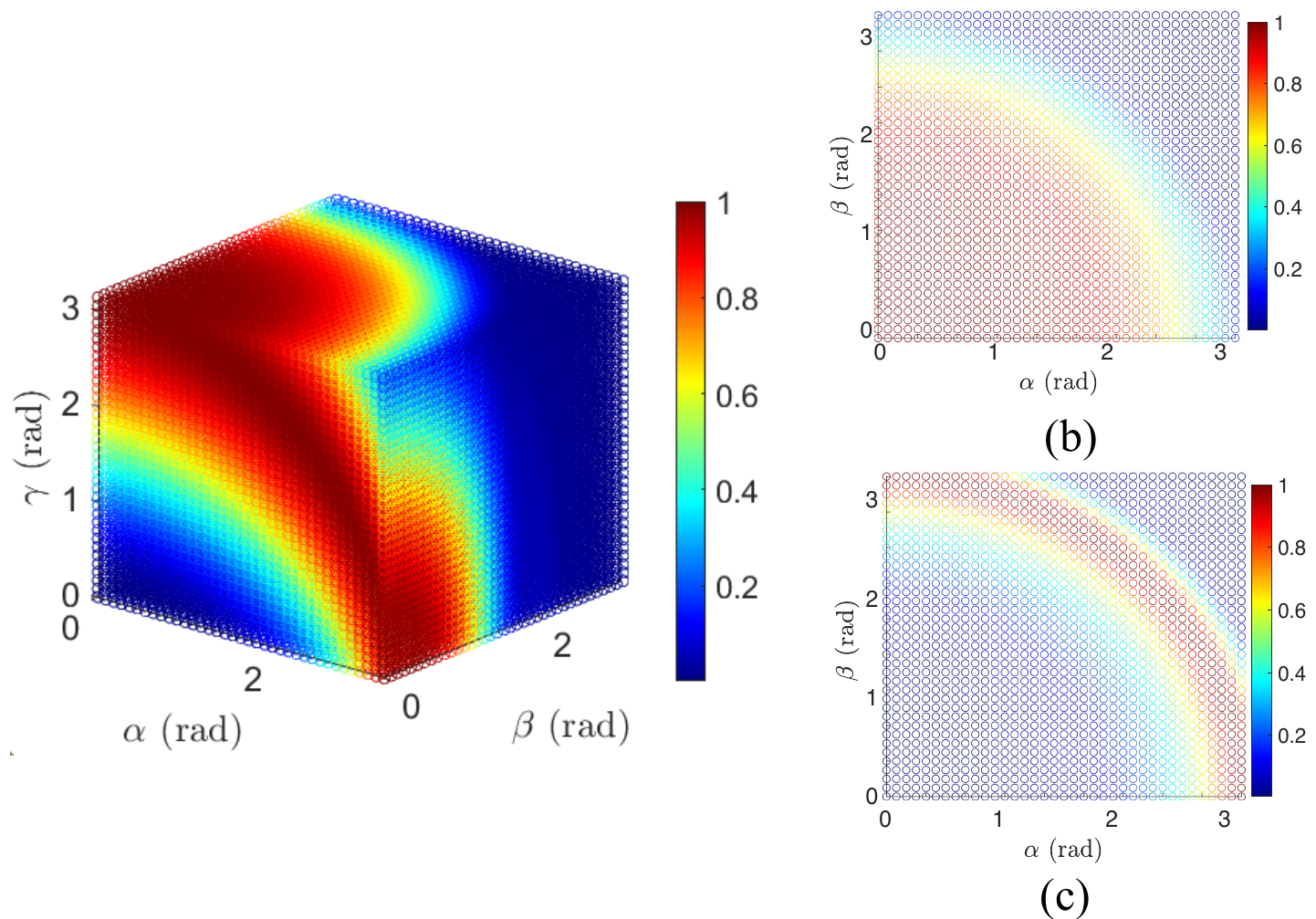}
    \vspace{-1ex}
    \caption{Feasible region for two layer 3D array structure.}
    \label{Feasible region analysis}
    \vspace{-1ex}
\end{figure}
\par Fig. \ref{Feasible region analysis} shows the spatial radiation characteristics of a 3D array with an interlayer spacing of $0.5\lambda$. Specifically, Fig. \ref{Feasible region analysis} (a) illustrates the normalized main beam intensity in space under different $(\alpha,\beta,\gamma)$ settings. From the figure, it is evident that for any given $\gamma$ value, the region with higher spatial radiation energy corresponds to an annular region in the $(\alpha,\beta)$ plane. This is consistent with our previous feasible region theories of the 3D array radiation characteristics. Additionally, Fig. \ref{Feasible region analysis} (b) and (c) show the $(\alpha,\beta,\gamma)$ plane for $\gamma = \pi$ and $\gamma=0$ settings, respectively. These figures demonstrate that the feasible region of the 3D array varies with different $\gamma$ settings, which needs to be considered in practical applications.
\begin{figure}[!ht]
    \vspace{0ex}
    \centering
    \includegraphics[width=3.3in]{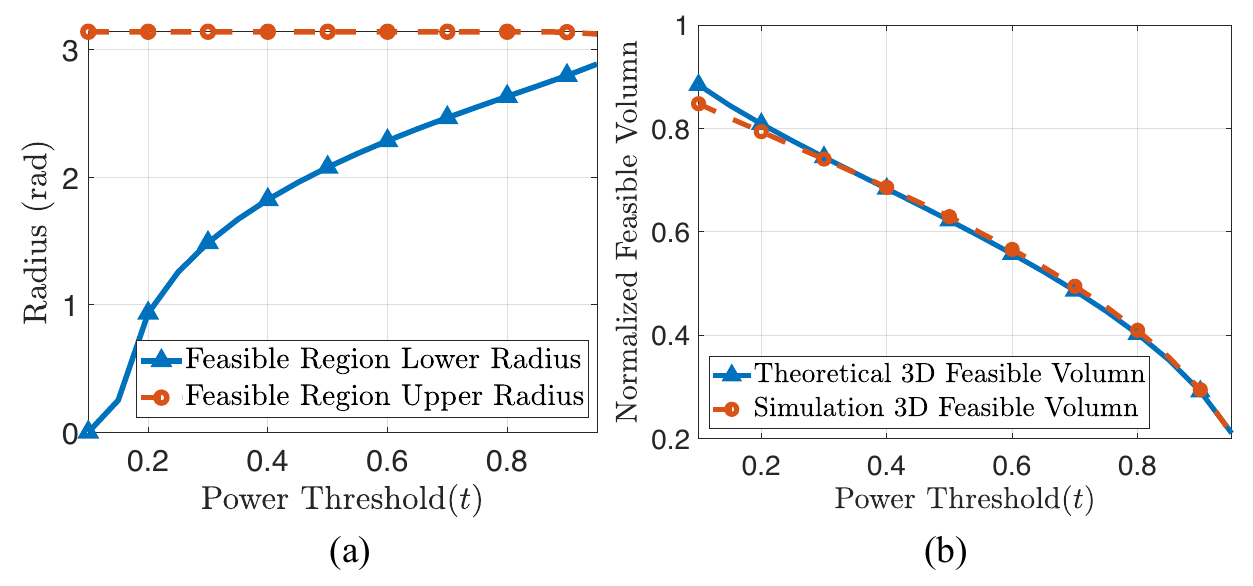}
    \vspace{-1ex}
    \caption{Feasible region for two layer 3D array structure.}
    \label{Temp6}
    \vspace{-0ex}
\end{figure}

Fig. \ref{Temp6} presents the simulation results of the feasible region analysis for the 3D antenna array. Fig. \ref{Temp6} (a) shows the upper and lower bounds of the feasible region annulus for a fixed beamforming direction of $\cos\xi = \frac{1}{4}$ under different energy attenuation thresholds. It can be observed that in the region with higher energy attenuation thresholds, the radius of the feasible region boundary is approximately linear with the energy attenuation threshold. Fig. \ref{Temp6} (b) displays the relationship between the feasible region volume and the energy attenuation threshold, where it can be seen that the theoretically calculated feasible region volume in (\ref{feasible volumn}) matches the simulated feasible region volume shown in Fig. \ref{Feasible region analysis} (a), verifying the correctness of the feasible region analysis theory.

\section{Conclusion}

\par In this paper, we extend the Hannan Limitation theory on the radiation efficiency of infinite 2D planar arrays to finite-dimensional 2D arrays by leveraging the symmetry of the DFT and the relationship between reflection coefficients and mutual coupling coefficients. 
Furthermore, we investigate the efficiency and gain limits of a two-layer 3D array structure to obtain the achievable spatial beamforming gain without increasing the planar aperture size. Specifically, results for both infinite and finite array cases are provided. 
Additionally, the spatial radiation and energy distribution characteristics of the considered two-layer 3D array structure are also investigated, presenting the feasible region of planar phase settings under a given energy attenuation threshold.
Through simulations, we demonstrate the validity of our proposed theories for both efficiency and gain limits. Additionally, we verify the efficacy of our proposed 3D array structure for better spatial coverage and the theoretical analysis of the spatial radiation characteristics of the 3D antenna array.

\bibliographystyle{IEEEtran}
\bibliography{IEEEabrv,references} 

\end{document}